\pdfoutput=1

\documentclass[a4paper,11pt]{article}

\usepackage{jcappub} 

\usepackage[T1]{fontenc} 
\usepackage{caption}
\usepackage{subcaption}
\usepackage{graphicx}
\usepackage{mathrsfs}
\usepackage{amsmath}
\usepackage{amssymb}
\usepackage{mathtools} 
\usepackage{soul}
\usepackage{bm}

\usepackage[utf8]{inputenc}

\usepackage{color}
\definecolor{amethyst}{rgb}{0.6, 0.4, 0.8}
\definecolor{alizarin}{rgb}{0.82, 0.1, 0.26}
\definecolor{green}{rgb}{0.55, 0.71, 0.0}
\definecolor{apricot}{rgb}{0.98, 0.81, 0.69}
\definecolor{auburn}{rgb}{0.43, 0.21, 0.1}
\definecolor{babyblueeyes}{rgb}{0.63, 0.79, 0.95}
\definecolor{bittersweet}{rgb}{1.0, 0.44, 0.37}
\definecolor{arsenic}{rgb}{0.23, 0.27, 0.29}
\usepackage{ulem}
\newcommand{\cfsout}{\bgroup\markoverwith{\textcolor{red}{\rule[0.5ex]{2pt}{0.4pt}}}\ULon}

\providecommand{\abs}[1]{\lvert#1\rvert}
\providecommand{\bd}[1]{\boldsymbol{#1}}
\providecommand{\ro}[1]{\mathrm{#1}}

\setstcolor{blue}
\sethlcolor{cyan}

\title{\boldmath Monopole Acceleration in Intergalactic Magnetic Fields}

\author[a,b,c]{Daniele Perri,}
\author[a,b,c]{Kyrilo Bondarenko,}
\author[d,e]{Michele Doro}
\author[a,b,c,f]{and Takeshi Kobayashi}

\affiliation[a]{SISSA, International School for Advanced Studies, \\Via Bonomea 265, 34136 Trieste, Italy\\}

\affiliation[b]{INFN, Sezione di Trieste, \\Via Valerio 2, 34127 Trieste, Italy\\}

\affiliation[c]{IFPU, Institute for Fundamental Physics of the Universe, \\Via Beirut 2, 34014 Trieste, Italy\\}

\affiliation[d]{University of Padova, Department of Physics and Astronomy, \\ I-35131, Padova, Italy\\}

\affiliation[e]{INFN, Sezione di Padova, \\Via Marzolo 8, 35131 Padova, Italy\\}

\affiliation[f]{Kobayashi-Maskawa Institute for the Origin of Particles and the Universe,\\ Nagoya University, Nagoya 464-8602, Japan\\}

\emailAdd{dperri@sissa.it}
\emailAdd{kyrylo.bondarenko@gmail.com}
\emailAdd{michele.doro@unipd.it}
\emailAdd{takeshi.kobayashi@sissa.it}

\abstract{
We provide a comprehensive analysis of the acceleration of magnetic monopoles in intergalactic magnetic fields. 
We demonstrate that monopoles with intermediate to low masses
can be accelerated to relativistic velocities. This can significantly affect direct and indirect searches for magnetic monopoles. As an example, we show that the Parker bound is relaxed in the presence of intergalactic fields. We also find that a cosmic population of monopoles can produce significant backreaction on the intergalactic fields.
}

\begin{document}
\maketitle
\flushbottom

\section{Introduction}
\label{sec:intro}

The study of magnetic monopoles (MMs) dates back to the pioneering work of Dirac~\cite{Dirac:1931kp}, who showed that MMs are consistent with quantum mechanics if their 
magnetic charges are quantized in units of $2 \pi / e$.
MMs would thus not only make the Maxwell's equations symmetric between electricity and magnetism, 
but would also be related to the observed quantization of electric charge. 
Later it was shown that many theories of particle physics, such as Grand Unified Theories, predict the existence of MMs as topological solitons \cite{tHooft:1974kcl,Polyakov:1974ek}, and that they can be created in the early universe during symmetry breaking phase transitions 
\cite{Zeldovich:1978wj,Preskill:1979zi}.
The mass of such solitonic monopoles is tied to the symmetry breaking scale, which depends on the details of the theory and thus can vary over many orders of magnitude.
While MMs have not been detected thus far, the search for MMs continues to be an active field of research with a potential to unlock new insights into the nature of our universe.

MMs are accelerated by the cosmic magnetic fields that exist in the universe on various scales -- from planets and stars to galaxies, galaxy clusters, filaments, and beyond.
Of particular relevance for a cosmic population of MMs
are the intergalactic magnetic fields (IGMFs) which permeate the immense intergalactic voids 
over cosmological distances. The IGMF has been for long constrained by upper limits on the intensity, mainly from cosmological observations of the cosmic microwave background as $B \lesssim 10^{-9}\, \ro{G}$. 
However in the past decade, the existence of IGMF has been suggested from 
gamma-ray experiments with the non-observation of extended halos or delayed emission around blazars formed by secondary photons
\cite{Tavecchio:2010mk,Neronov:2010gir,Dermer:2010mm,MAGIC:2022piy}.
These studies set \textit{lower limits} on the IGMF strength of
$B \gtrsim 10^{-15}\, \ro{G}$,
if the correlation length~$\lambda$ is of Mpc scale or larger.
If $\lambda$ is much smaller than a Mpc, the lower limit further improves as $\lambda^{-1/2}$.
See e.g.~\cite{Durrer:2013pga,AlvesBatista:2021sln} for reviews on IGMFs.\footnote{A non-IGMF explanation of the blazar observations that invokes a beam plasma instability was also proposed~\cite{Broderick:2011av,Miniati:2012ge}, although this mechanism has been questioned in, e.g.,~\cite{Perry:2021rgv,Alawashra:2022all}.}

In light of the observational results, in this paper we analyze in detail the acceleration of MMs in IGMFs. Although IGMFs are extremely weak compared with Galactic fields which are of $B \sim 10^{-6}\, \ro{G}$, their very large coherence lengths contribute significantly to the MM acceleration.
We demonstrate that IGMFs can easily accelerate the MMs to relativistic velocities. 
(See also the earlier works \cite{Kephart:1995bi,Escobar:1997mr,Wick:2000yc}
which discussed relativistic MMs from Galactic/intergalactic fields.)
Moreover, we show that the backreaction of the MMs on the IGMFs can drastically affect the acceleration process and must be taken into account for an accurate calculation of the MM velocity. 
As we are interested in the average velocity of the cosmic population of MMs, we mainly focus on the universe-filling IGMFs. 
There can be additional extragalactic magnetic fields, for instance fields in cosmic filaments, as well as fields transported by galactic winds 
(see, e.g. \cite{Akahori:2017lhe, Garcia:2020kxm, Erceg_2022, Carretti:2022fqk, OSullivan:2023eub, Heesen_2023}).
However it is expected that such fields do not fill the entire universe~\cite{Bertone:2006mr}, and thus can only affect the velocity of a small fraction of the cosmic MMs. We hence exclude them from our analysis.

Cosmic  magnetic fields lose energy when they accelerate MMs. Therefore, the requirement for the survival of the fields imposes an upper limit on the abundance of MMs. This idea was first introduced by Parker, who calculated an upper limit on the MM flux within our Galaxy based on the survival of Galactic magnetic fields \cite{Parker:1970xv, Turner:1982ag}. 
This limit, known as the Parker bound, was later extended to the survival of the seed magnetic field of our Galaxy \cite{Adams:1993fj},
as well as magnetic fields in galactic clusters \cite{Rephaeli:1982nv}. 
Additionally, if IGMFs have a primordial origin, as suggested by numerous studies (see \cite{Grasso:2000wj,Subramanian:2015lua} for reviews), their survival also leads to Parker-type constraints \cite{Long:2015cza, Kobayashi:2022qpl,Kobayashi:2023ryr}. 
Here we note that in literature, Parker bounds based on Galactic magnetic fields have been derived under the assumption that MMs have a velocity of $v \sim 10^{-3}$, which corresponds to the virial velocity or the peculiar velocity of the Milky Way Galaxy~\cite{Kogut:1993ag}. 
However, the bounds strongly depend on the kinetic energy of the MMs,
and thus are significantly affected by processes that accelerate MMs before they enter the Galaxy.
As an application of our study of the MM acceleration in IGMFs, we show how Galactic Parker bounds are modified by taking into account the acceleration effects. In particular, we find that the bounds from the survival of seed Galactic magnetic fields can be significantly relaxed in the presence of IGMFs within the observational limits.

This paper is organized as follows. 
In Section~\ref{sec:acc} we analyze MM acceleration in IGMFs, and the backreaction on the IGMFs.
In Section~\ref{sec:Parker} we show how Galactic Parker bounds are affected by the MM acceleration in IGMFs. 
We then conclude in Section~\ref{sec:concl}. 
In Appendix~\ref{app:acceleration}, we consider the hypothesis that the IGMFs have a primordial origin and show that MM acceleration in primordial fields does not modify the order-of-magnitude results of the main text. 
In Appendix~\ref{app:energy_loss_medium} we evaluate the energy loss of fast-moving MMs in the intergalactic space and inside the Galaxy.
In Appendix~\ref{app:oscillation} we study energy oscillations between MMs and magnetic fields.
In Appendix~\ref{app:MonAccTot} we compute the MM velocity at Earth by further taking into account the acceleration in Galactic magnetic fields.

Throughout this work we use Heaviside-Lorentz units, with $c = \hbar = k_B = 1$.
We denote the MM mass by~$m$, and the amplitude of the magnetic charge by~$g$.
The fundamental Dirac charge of a MM is written as $g_{\mathrm{D}} = 2 \pi / e \approx 21$.

\section{Acceleration of monopoles in intergalactic magnetic fields}
\label{sec:acc}

In this section, we derive the velocity of MMs after they have been accelerated in IGMFs for a Hubble time. 
We start by treating the IGMFs as a background. Then we discuss the backreaction of MMs on the IGMFs, and evaluate the actual velocity that MMs obtain in the intergalactic space.

\subsection{Acceleration in an intergalactic magnetic field background}

Let us model IGMFs with coherence length~$\lambda_{\mathrm{I}}$, by dividing the universe into cells of uniform field, with each cell having a size~$\lambda_{\mathrm{I}}$.
We take the field strength in all cells to have the same value~$B_{\ro{I}}$, and compute the MM velocity after a Hubble time. 
We will occasionally ignore numerical factors of order unity.\footnote{The final results of this paper are only mildly sensitive to the ignored factors, such that the results give the correct order of magnitude (unless there is a drastic modification to the magnetic field model itself.) The plots can also be trusted at the order-of-magnitude level.}

\subsubsection{First cell}

Let us consider all MMs to be initially at rest at $t = 0$.
They are then accelerated by a uniform field in the first cell as
\begin{equation}
 m \gamma v = g B_{\mathrm{I}} t,
\label{eq:B1}
\end{equation}
where $\gamma = 1 / \sqrt{1 - v^2}$.
This can be used to compute the time it takes for the MMs to travel a distance $\sim \lambda_{\mathrm{I}}$ and exit the first cell as
\begin{equation}
 \Delta t_1 \sim  
\ro{max.} \left\{
\lambda_{\mathrm{I}} , \,  \left( \frac{m \lambda_{\mathrm{I}} }{g B_{\mathrm{I}} }\right)^{1/2}
\right\}.
\label{eq:tau_1}
\end{equation}
The first term applies to MMs that are accelerated to relativistic velocities within the first cell, while the second term is for MMs that stay nonrelativistic.

\subsubsection{Second cell onward}

As the MMs pass through multiple cells, they are deflected by the magnetic field in each cell.
Assuming the directions of the field to be uncorrelated from one cell to the next, 
the average kinetic energy of each MM grows with the number of cells crossed~$N$ as\footnote{See \cite{Kobayashi:2023ryr} for a detailed derivation. The second line of their Eq.~(A.15) applies to our current case with a vanishing initial velocity.}
\begin{equation}
 m (\gamma - 1 )\sim  g B_{\mathrm{I}} \lambda_{\mathrm{I}}
N^{1/2}.
\label{eq:B3}
\end{equation}
The number of cells each MM crosses before becoming relativistic can be read off from (\ref{eq:B3}) as,
\begin{equation}
 N_{\ro{rel}} \sim \left( \frac{m}{g B_{\mathrm{I}} \lambda_{\mathrm{I}}}  \right)^2.
\end{equation}
Using this, (\ref{eq:B3}) can be rewritten as an expression for the product~$\gamma v$ in the nonrelativistic and relativistic regimes as,
\begin{equation}
\gamma v \sim 
  \begin{dcases}
\left( \frac{N}{N_{\ro{rel}}} \right)^{1/4}
  & \mathrm{for}\, \, \,  
N < N_{\ro{rel}},
 \\
\left( \frac{N}{N_{\ro{rel}}} \right)^{1/2}
  & \mathrm{for}\, \, \,  
N > N_{\ro{rel}}.
 \end{dcases}
\label{eq:B5}
\end{equation}
If $m < g B_{\mathrm{I}} \lambda_{\mathrm{I}}$ (i.e. $N_{\ro{rel}} < 1$), the MMs become relativistic within the first cell and thus the second line in (\ref{eq:B5}) holds for all~$N$.

One sees from (\ref{eq:B5}) that from the second cell onward, the MM velocity does not change in each cell by more than an order-unity factor. Hence the crossing time for the $N$th cell can be estimated as
\begin{equation}
 \Delta t_N \sim \frac{\lambda_{\mathrm{I}}}{v},
\end{equation}
using the exit velocity~$v$ from the $N$th cell as given in (\ref{eq:B5}). 
For $2 \leq N < N_{\ro{rel}}$, the first line of (\ref{eq:B5}) yields
\begin{equation}
\Delta t_N  \sim 
\lambda_{\mathrm{I}} \left( \frac{N_{\ro{rel}}}{N} \right)^{1/4}.
\label{eq:tau_N-non}
\end{equation}
At $N > N_{\ro{rel}}$, the MM is relativistic and hence the crossing time becomes 
\begin{equation}
 \Delta t_N \sim \lambda_{\mathrm{I}}.
\label{eq:tau_N-rel}
\end{equation}
These expressions for the cell-crossing time can also be used for the first cell ($N=1$), since (\ref{eq:tau_N-non}) and (\ref{eq:tau_N-rel}) 
match respectively with the second and first terms in (\ref{eq:tau_1}).
By adding up (\ref{eq:tau_N-non}) and (\ref{eq:tau_N-rel}) for all cells, 
and using the approximation $\sum_{n = 1}^N n^{-1/4} \sim (4/3) N^{3/4}$,
one can express the elapsed time in terms of the number of crossed cells as,
\begin{equation}
t = \sum_{n=1}^N \Delta t_n \sim
  \begin{dcases}
\lambda_{\mathrm{I}} N_{\ro{rel}}^{1/4} N^{3/4} 
  & \mathrm{for}\, \, \,  
N < N_{\ro{rel}},
 \\
\lambda_{\mathrm{I}} N
  & \mathrm{for}\, \, \,  
N > N_{\ro{rel}}.
\label{eq:B12}
 \end{dcases}
\end{equation}
One sees from this result that MMs with $m > g B_{\mathrm{I}} \lambda_{\mathrm{I}}$ become relativistic after a time period of
$t_{\ro{rel}} \sim \lambda_{\mathrm{I}} N_{\ro{rel}}$.

\subsubsection{Velocity after a Hubble time}

We are now ready to evaluate the velocity of monopoles that have been accelerated in an IGMF background.
We consider the acceleration over a Hubble time~$1/H_0$, and hence ignore cosmic expansion. In Appendix~\ref{app:acceleration} we show that, 
even if the IGMF are remnants of primordial fields and the MMs have been accelerated since the early universe, our estimate of the final MM velocity is modified only by order-unity factors. 

\textit{Homogeneous IGMF.}
If the magnetic coherence length is larger than the Hubble radius, 
$\lambda_{\mathrm{I}} > 1/ H_0 $, 
then the field is effectively homogeneous and the final value of $\gamma v$ is obtained by substituting $t = 1/H_0$ into (\ref{eq:B1}).
This yields
\begin{equation}
\label{eq:homo}
    \left( \gamma v \right)_0 \sim \frac{g B_{\mathrm{I}}}{m H_0} .
\end{equation}

\textit{Inhomogeneous IGMF.}
With sub-horizon coherence lengths, $\lambda_{\mathrm{I}} < 1/ H_0 $, 
the present-day velocity takes the forms,
\begin{equation}
\left( \gamma v \right)_0 \sim
  \begin{dcases}
\frac{gB_{\mathrm{I}} \lambda_{I}}{m }
\frac{1}{(\lambda_{\mathrm{I}} H_0)^{1/2}}
  & \mathrm{for}\, \, \,  
m < \frac{g B_{\mathrm{I}} \lambda_{\mathrm{I}}^{1/2}}{H_0^{1/2}},
 \\
\left( \frac{gB_{\mathrm{I}} \lambda_{\mathrm{I}}}{m } \right)^{2/3}
\frac{1}{(\lambda_{I} H_0 )^{1/3}}
  & \mathrm{for}\, \, \,  
\frac{g B_{\mathrm{I}}  \lambda_{\mathrm{I}}^{1/2}}{H_0^{1/2}} < m < 
\frac{g B_{\mathrm{I}}}{\lambda_{\mathrm{I}} H_0^2} ,
 \\
\frac{gB_{\mathrm{I}}}{m H_0}
  & \mathrm{for}\, \, \,  
m > \frac{g B_{\mathrm{I}}}{\lambda_{\mathrm{I}} H_0^2}.
 \end{dcases}
\label{eq:nonhomoGen}
\end{equation}
These can be understood as follows.
Firstly, if the MM mass is as large as
$m > g B_{\mathrm{I}}/ \lambda_{\mathrm{I}} H_0^2$, 
then one sees from (\ref{eq:tau_1}) that $ \Delta t_1 > 1/H_0$.
The MMs thus do not exit their first cells,
and the final velocity is given by the same expression as~(\ref{eq:homo}). 
Lighter MMs, on the other hand, pass through multiple cells.
However if $\Delta t_1 < 1/H_0 < \lambda_{\mathrm{I}} N_{\ro{rel}}$, then the MMs on average stay nonrelativistic after a Hubble time. 
Hence the final velocity can be obtained by combining the first lines of (\ref{eq:B5}) and (\ref{eq:B12}) with $t = 1/H_0$;
this yields the second line of~(\ref{eq:nonhomoGen}).
Finally, for even lighter MMs satisfying $1/H_0 > \lambda_{\mathrm{I}} N_{\ro{rel}}$, the second lines of 
(\ref{eq:B5}) and (\ref{eq:B12}) yield the relativistic velocity
in the first line of~(\ref{eq:nonhomoGen}).

In the above discussions we have ignored the energy loss of MMs due to interactions with the intergalactic medium and radiative emissions. 
In Appendix~\ref{app:energy_loss_medium} we show that these effects are actually negligible.

\subsection{Backreaction on intergalactic magnetic fields}

We have thus far treated the IGMFs as a background. 
However the MMs extract energy from the IGMFs as they are accelerated, and hence the total kinetic energy of the MMs cannot become larger than the initial energy of the IGMFs.
We therefore have a constraint,
\begin{equation}
 n m (\gamma - 1) < \frac{B_{\mathrm{I}}^2}{2} ,
\label{eq:maru-i}
\end{equation}
where $n$ is the average number density of the MMs in the universe. 
This sets the maximal Lorentz factor of the MMs,
\begin{equation}
 \gamma_{\ro{max}} - 1 = \frac{B_{\mathrm{I}}^2}{2n m },
\label{eq:maru-ii}
\end{equation}
from which one can also obtain the maximal velocity~$v_{\ro{max}}$.

If the velocity~$v_0$ given in Eqs.~(\ref{eq:homo}) or (\ref{eq:nonhomoGen}) is much smaller than~$v_{\ro{max}}$, 
the backreaction on the IGMFs is negligible.
On the other hand if the MM velocity approaches~$v_{\ro{max}}$, then it implies that the energy of the IGMFs has been transferred to the MMs. 
However the MMs do not effectively dissipate the gained energy into the intergalactic medium as shown in Appendix~\ref{app:energy_loss_medium}, 
and thus they eventually return the energy to the IGMFs.
This initiates an energy oscillation between the IGMFs and the population of MMs~\cite{Long:2015cza},
with the oscillation-averaged MM velocity being of order~$v_{\ro{max}}$.
We expect the IGMF-MM oscillation to avoid Landau damping; see Appendix~\ref{app:oscillation} for discussions on this point.

The MM velocity in the above two cases can collectively be written as
\begin{equation}
 v_{\ro{CMB}} = \ro{min.} \left\{ v_0, v_{\ro{max}} \right\} .
\label{eq:v_CMB}
\end{equation}
Here we used the subscript ``CMB'' to highlight that this is the MM velocity with respect to the CMB rest frame. We also write the MM flux 
in the CMB rest frame per area per time per solid angle\footnote{This expression for the flux implicitly assumes the MMs to be moving in random directions, which is not the case if the IGMF is homogeneous.
However this is good enough for obtaining order-of-magnitude results.} as 
$F_{\ro{CMB}} = n v_{\ro{CMB}} / 4 \pi$.

\begin{figure}[h!t!]
    \includegraphics[width=0.65\textwidth]{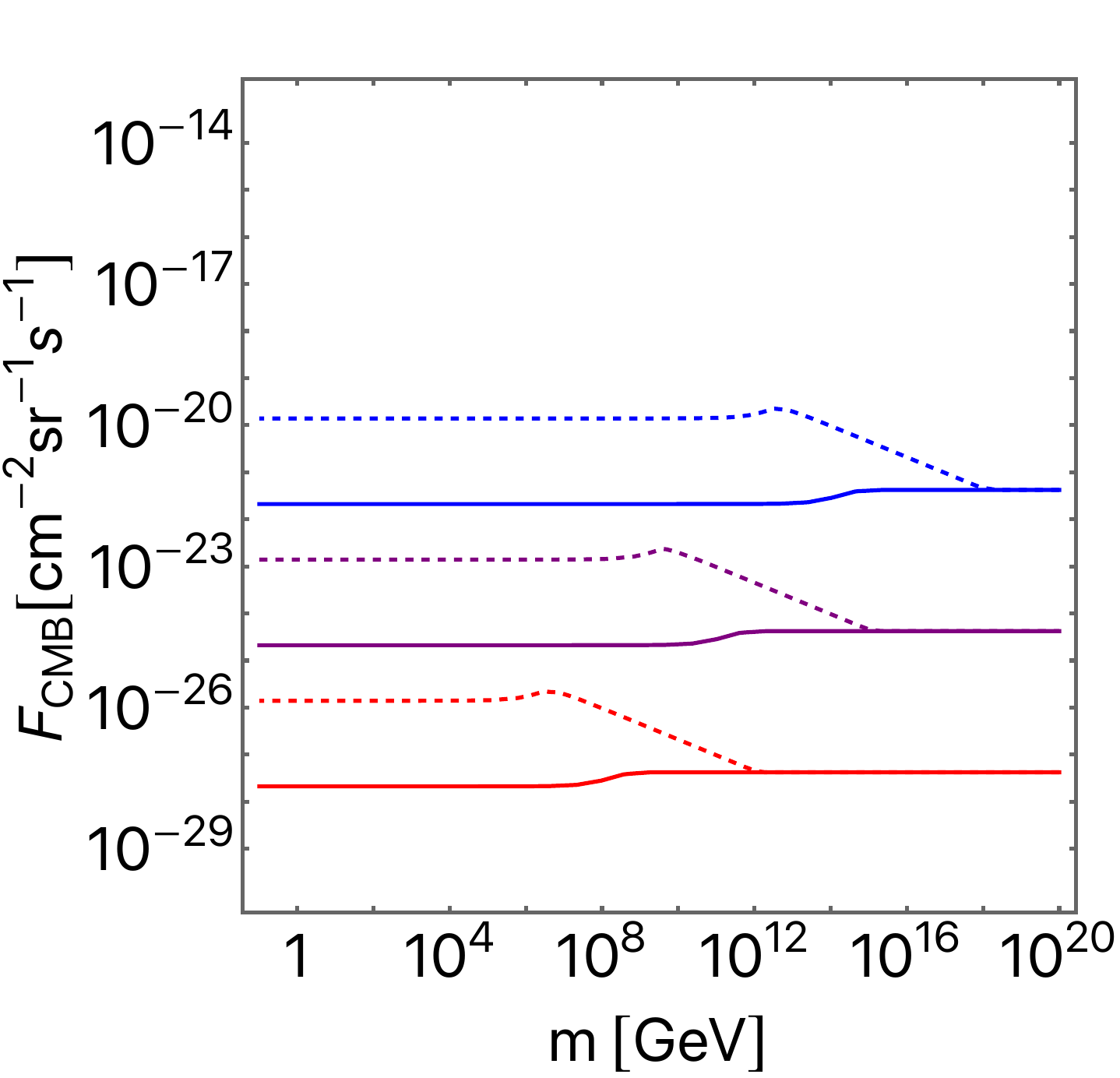}
    \centering
    \caption{Threshold value of the MM flux beyond which the backreaction to the IGMFs becomes significant. 
The IGMF amplitude is varied as
$B_{\mathrm{I}} = 10^{-9}\, \mathrm{G}$ (blue), 
$ 10^{-12}\, \mathrm{G}$ (purple),
$10^{-15}\, \mathrm{G}$ (red),
with the correlation length
$\lambda_{\mathrm{I}} > 1/H_0$ (solid) and $\lambda_{\mathrm{I}} = 1\, \mathrm{Mpc}$ (dashed).
The MM charge is fixed to $g=g_{\mathrm{D}}$.}
    \label{fig:feq}
\end{figure}

In the case where $v_0 > v_{\ro{max}}$, 
the expression~(\ref{eq:maru-ii}) for the maximal velocity can be rewritten in terms of the flux as
\begin{equation}
\label{eq:gammaMax}
    \gamma_{\ro{max}} - 1 = \frac{B_{\mathrm{I}}^2 v_{\ro{max}}}{8 \pi m F_{\ro{CMB}}} .
\end{equation}
This can be solved in the nonrelativistic and ultrarelativistic limits to give the product~$(\gamma v)_{\ro{max}}$ as,
\begin{equation}\label{eq:gamma_v_max}
(\gamma v)_{\ro{max}} \simeq
  \begin{dcases}
\frac{B_{\mathrm{I}}^2}{4 \pi m F_{\ro{CMB}}}
  & \mathrm{for}\, \, \,  
B_{\mathrm{I}}^2 \ll 8 \pi m F_{\ro{CMB}},
 \\
\frac{B_{\mathrm{I}}^2}{8 \pi m F_{\ro{CMB}}}
  & \mathrm{for}\, \, \,  
B_{\mathrm{I}}^2 \gg 8 \pi m F_{\ro{CMB}}.
 \end{dcases}
\end{equation}
Notice that $v_{\ro{max}}$ is independent of the IGMF coherence length, while it depends on the MM flux (or the density).

\begin{figure*}[h!t]
     \centering
     \hspace{35pt}
     \begin{subfigure}[b]{0.55\textwidth}
         \centering
         \includegraphics[width=0.8\textwidth]{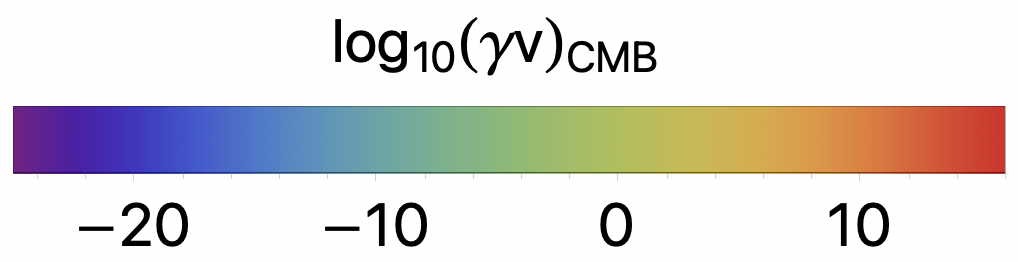}
     \end{subfigure}
     \hspace{65pt}
     \begin{subfigure}[b]{0.49\textwidth}
         \centering
         \includegraphics[width=\textwidth]{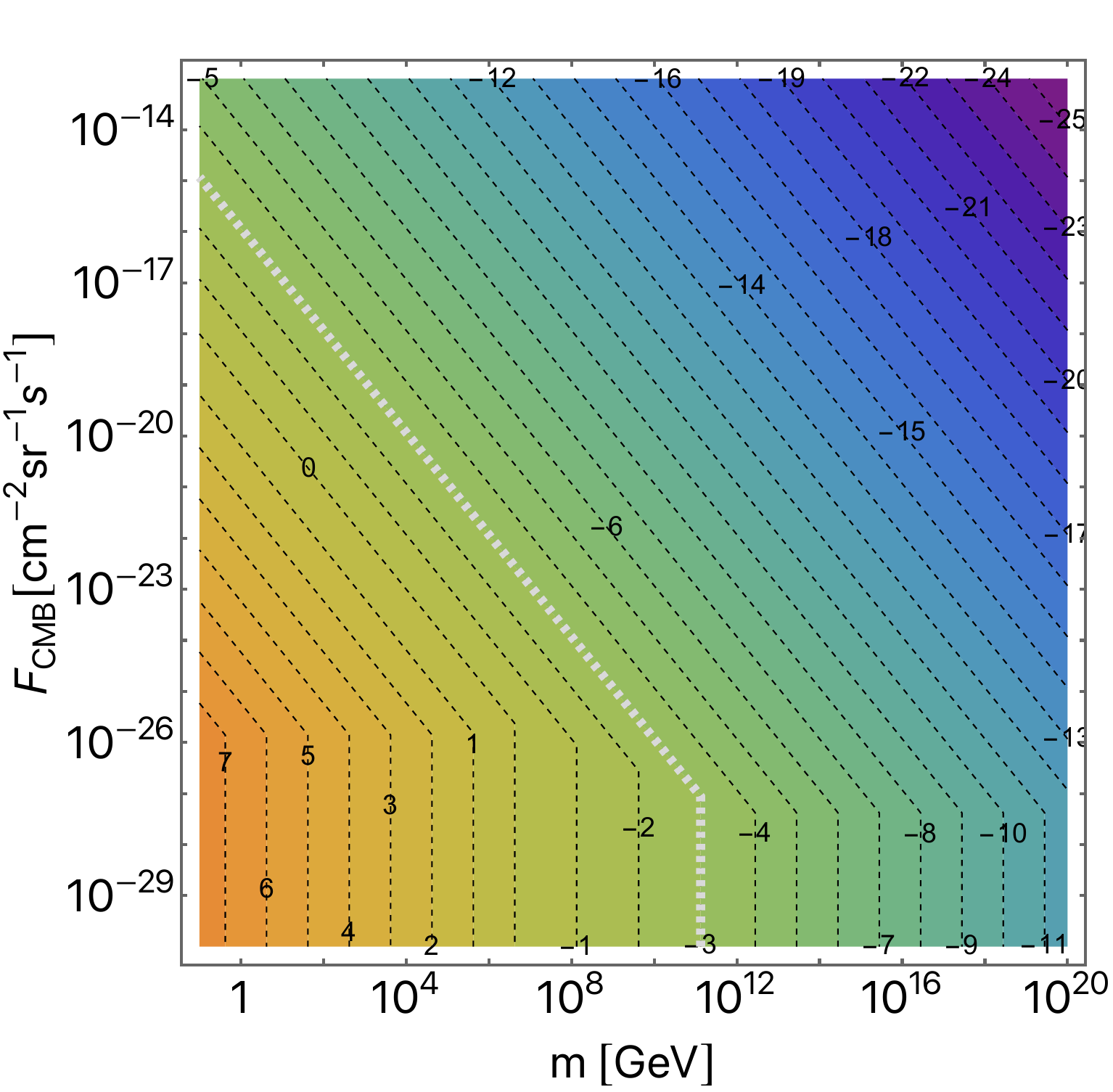}
         \caption{$B_{\mathrm{I}} = 10^{-15}\, \ro{G}$, $\lambda_{\mathrm{I}} = 1\, \mathrm{Mpc}$.}
         \label{fig:cmb15mpc}
     \end{subfigure}
     \hfill
     \begin{subfigure}[b]{0.49\textwidth}
         \centering
         \includegraphics[width=\textwidth]{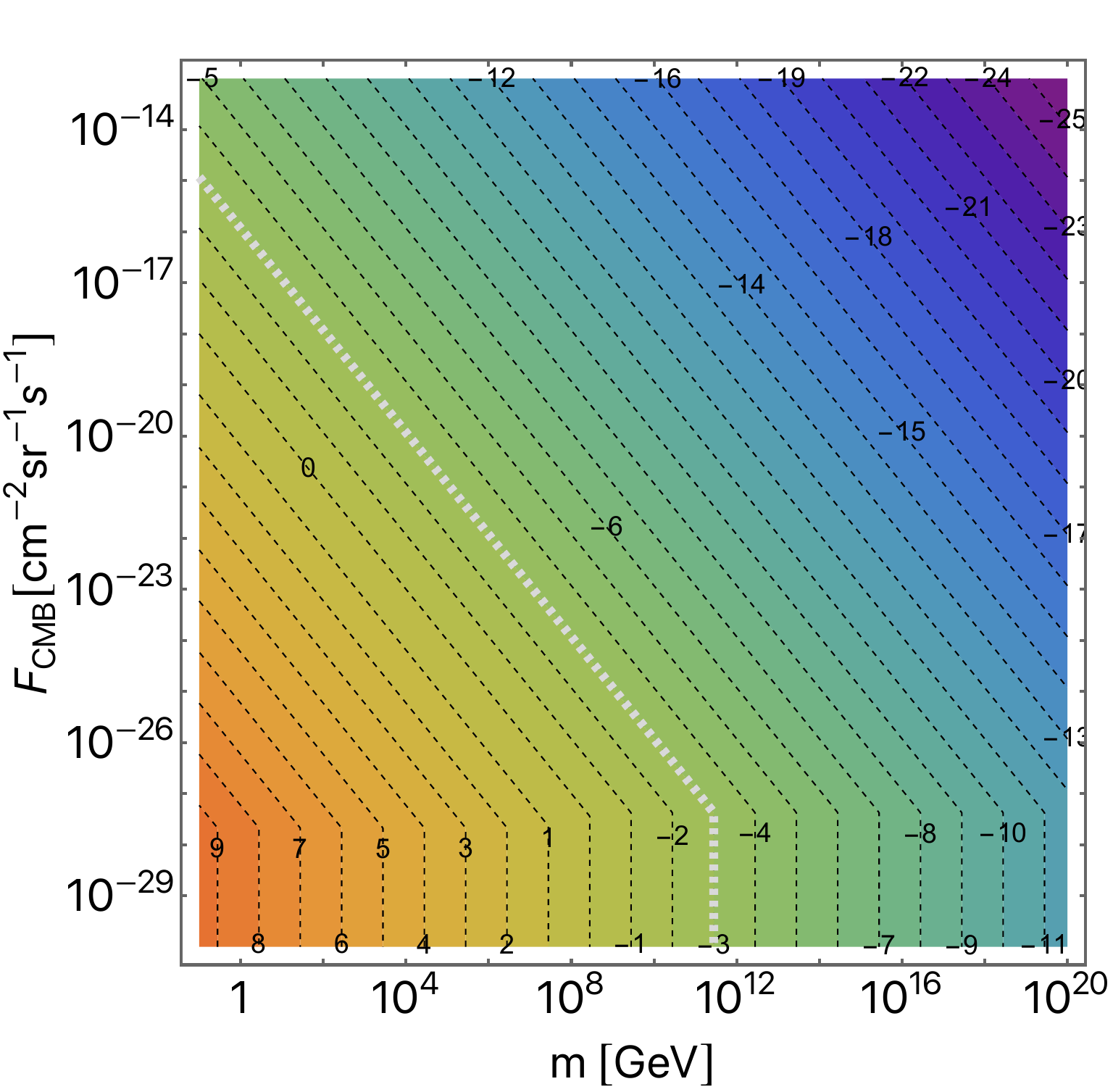}
         \caption{$B_{\mathrm{I}} = 10^{-15}\, G$, $\lambda_{\mathrm{I}} > 1/H_0$.}
         \label{fig:cmb15homo}
     \end{subfigure}
    \begin{subfigure}[b]{0.49\textwidth}
         \centering
         \includegraphics[width=\textwidth]{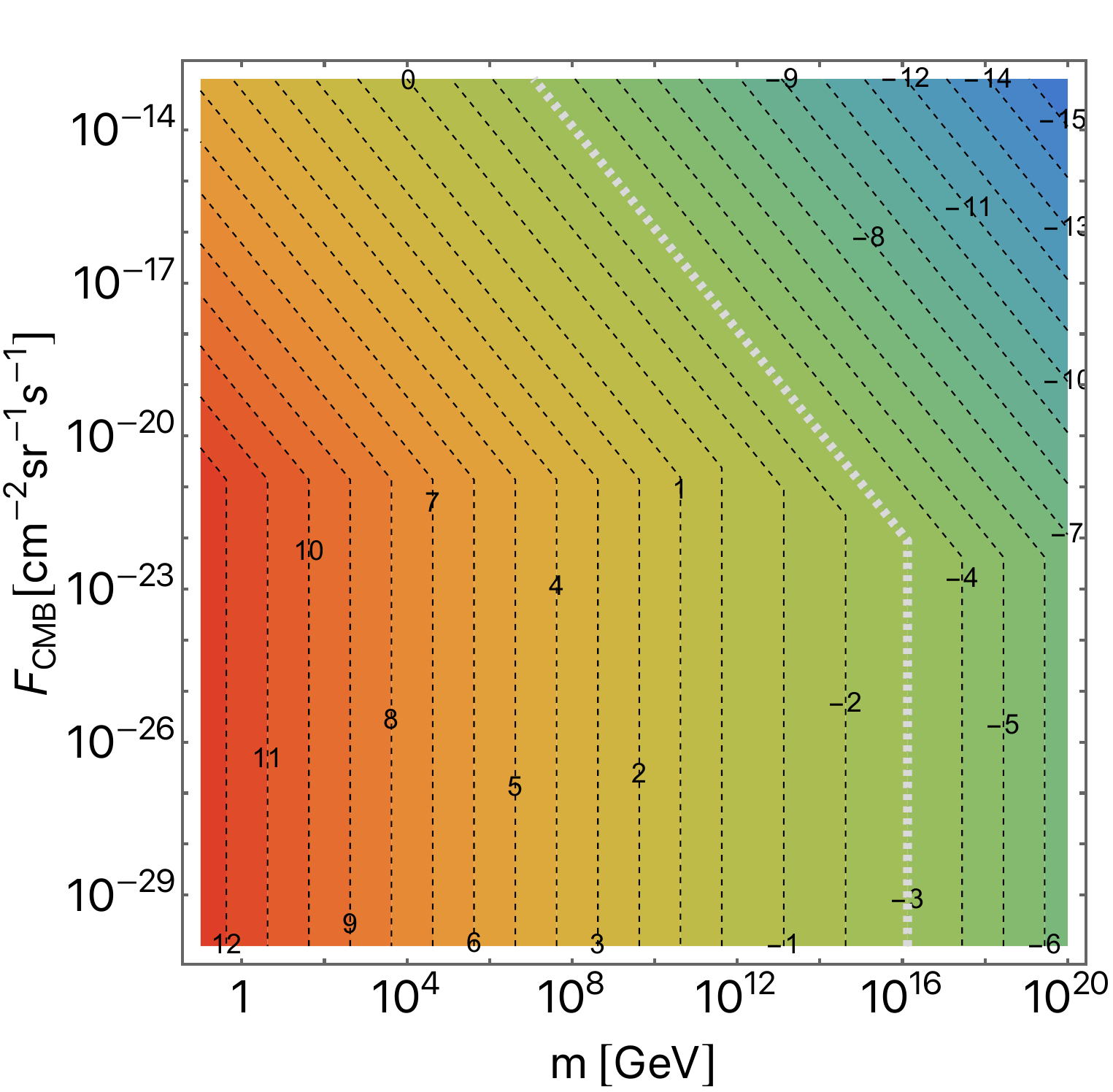}
         \caption{$B_{\mathrm{I}} = 10^{-10}\, G$, $\lambda_{\mathrm{I}} = 1\, \mathrm{Mpc}$.}
         \label{fig:cmb10mpc}
     \end{subfigure}
     \hfill
     \begin{subfigure}[b]{0.49\textwidth}
         \centering
         \includegraphics[width=\textwidth]{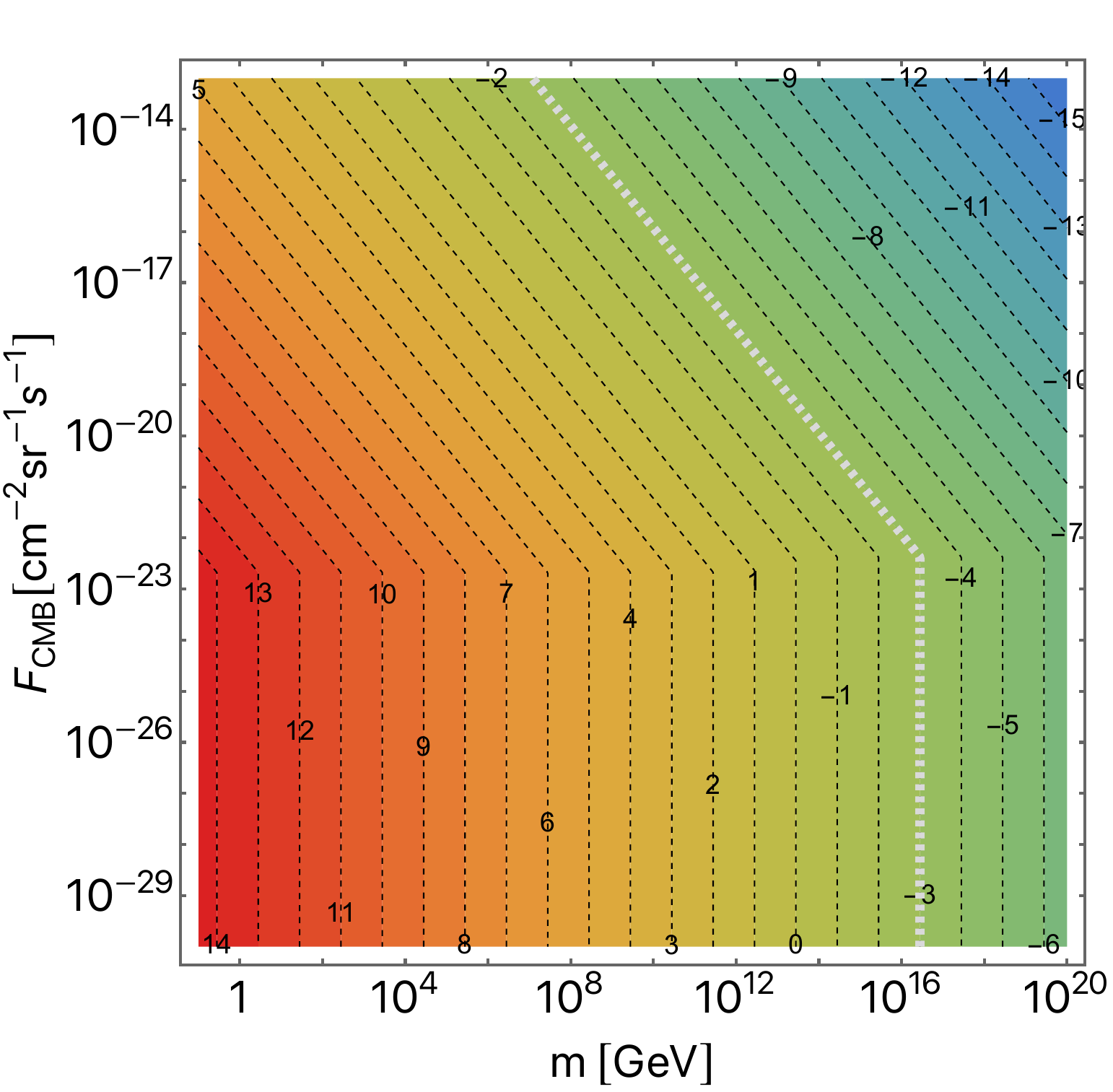}
         \caption{$B_{\mathrm{I}} = 10^{-10}\, G$, $\lambda_{\mathrm{I}} > 1/H_0$.}
         \label{fig:cmb10homo}
     \end{subfigure}
     \caption{Velocity of MMs accelerated in IGMFs, in the CMB rest frame. The contours show $\log_{10} ( \gamma v )_{\mathrm{CMB}}$.
The MM charge is fixed to $g = g_{\mathrm{D}}$, while the IGMF amplitude and coherence length are varied in each panel.
Dashed gray lines highlight where $v_{\mathrm{CMB}} = 10^{-3}$.}
        \label{fig:MonAccCMB}
\end{figure*}

Writing the threshold value of the MM flux beyond which the backreaction to the IGMFs becomes significant as~$F_{\ro{BR}}$, this can be obtained by solving $v_0 = v_{\ro{max}}$. 
For $\lambda_{\mathrm{I}} > 1/H_0$, matching Eqs.~\eqref{eq:homo} and~\eqref{eq:gamma_v_max} yields
\begin{equation}
\label{eq:feq_homo}
    F_{\mathrm{BR}} \sim \frac{B_{\mathrm{I}} H_0}{4 \pi g} .
\end{equation}
For $\lambda_{\mathrm{I}} < 1/H_0$, Eqs.~\eqref{eq:nonhomoGen} and~\eqref{eq:gamma_v_max} give
\begin{equation}
\label{eq:feq_nonhomo}
F_{\mathrm{BR}} \sim 
  \begin{dcases}
\frac{B_{\mathrm{I}}}{4 \pi g} \left( \frac{H_0}{\lambda_{\mathrm{I}}} \right)^{1/2}
  & \mathrm{for}\, \, \,  
m < \frac{g B_{\mathrm{I}}  \lambda_{\mathrm{I}}^{1/2}}{H_0^{1/2}},
 \\
\frac{1}{4 \pi} \left( \frac{B_{\mathrm{I}}^4 H_0}{g^2 m \lambda_{\mathrm{I}}} \right)^{1/3}
  & \mathrm{for}\, \, \,  
\frac{g B_{\mathrm{I}}  \lambda_{\mathrm{I}}^{1/2}}{H_0^{1/2}} < m < 
\frac{g B_{\mathrm{I}}}{\lambda_{\mathrm{I}} H_0^2} ,
 \\
\frac{B_{\mathrm{I}} H_0}{4 \pi g}
  & \mathrm{for}\, \, \,  
m > \frac{g B_{\mathrm{I}}}{\lambda_{\mathrm{I}} H_0^2} .
 \end{dcases}
\end{equation}
We note that since the two expressions in \eqref{eq:gamma_v_max} differ only by a factor~$2$, here we just used the first line to obtain order-of-magnitude estimates of~$F_{\ro{BR}}$.

In Figure~\ref{fig:feq} we plot $F_{\mathrm{BR}}$ as a function of the MM mass. 
The IGMF amplitude is varied as
$B_{\mathrm{I}} = 10^{-9}\, \mathrm{G}$ (blue), 
$ 10^{-12}\, \mathrm{G}$ (purple),
$10^{-15}\, \mathrm{G}$ (red),
with the correlation length taken as
$\lambda_{\mathrm{I}} > 1/H_0$ (solid lines) and $\lambda_{\mathrm{I}} = 1\, \mathrm{Mpc}$ (dashed lines).
The solid and dashed lines with the same color join at large masses.
The charge is fixed to $g=g_{\mathrm{D}}$.
We show MM masses down to $1 \, \mathrm{GeV}$, however we remark that this value is a few orders of magnitude smaller than the latest experimental lower limit~\cite{MoEDAL:2021vix}.
As one goes toward smaller masses each curve drops by a factor~$2$; this 
corresponds to the difference between the nonrelativistic and ultrarelativistic regimes as shown in Eq.~(\ref{eq:gamma_v_max}). 
The displayed results match with the approximate expressions \eqref{eq:feq_homo} and \eqref{eq:feq_nonhomo}
up to order-unity factors.
One sees that $F_{\ro{BR}}$ increases for larger~$B_{\ro{I}}$, since the IGMF becomes more resilient to the backreaction from the MMs. 
$F_{\ro{BR}}$ at small masses increases also for smaller~$\lambda_{\ro{I}}$, since the MM acceleration becomes less effective.

\begin{figure}[t]
    \includegraphics[width=0.65\textwidth]{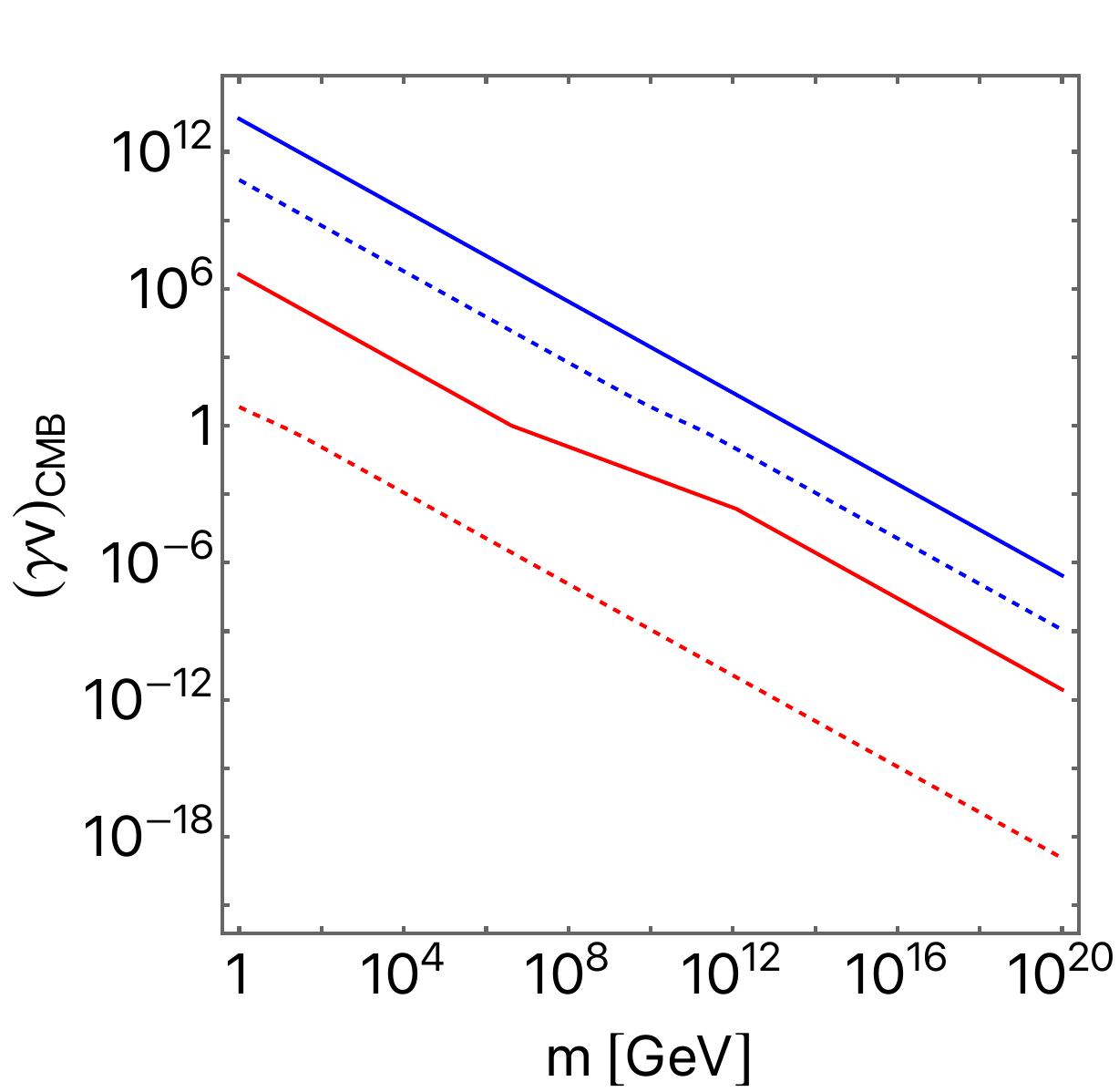}
    \centering
    \caption{MM velocity in the CMB rest frame for fixed values of the MM flux: 
$F_{\mathrm{CMB}} = 10^{-30}\, \mathrm{cm}^{-2} \, \mathrm{sr}^{-1}\,  \mathrm{s}^{-1}$ (solid lines) and
$10^{-20}\, \mathrm{cm}^{-2} \, \mathrm{sr}^{-1} \, \mathrm{s}^{-1}$ (dashed).
The IGMF strength and coherence length are taken as 
$B_{\mathrm{I}} = 10^{-15}\, \mathrm{G}$
and $\lambda_{\mathrm{I}}=1\, \mathrm{Mpc}$ for the red lines,
while $B_{\mathrm{I}} = 10^{-10}\, \mathrm{G}$
and $\lambda_{\mathrm{I}} > 1/H_0$ for the blue lines.
The MM charge is fixed to $g = g_{\mathrm{D}}$.}
    \label{fig:MonAcc}
\end{figure}

\subsection{Monopole velocity in the CMB rest frame}

In Figure~\ref{fig:MonAccCMB} we show the value of $(\gamma v)_{\ro{CMB}}$ in Eq.~\eqref{eq:v_CMB} as a function of the MM mass and flux in the CMB rest frame. 
We display four combinations of the IGMF strength ($B_{\mathrm{I}} = 10^{-15}\, \mathrm{G}$ and $10^{-10}\, \mathrm{G}$)
and correlation length 
($\lambda_{\mathrm{I}} = 1\, \mathrm{Mpc}$ and $\lambda_{\mathrm{I}} > 1/H_0$).
The MM charge is fixed to $g = g_{\mathrm{D}}$.
The values of~$F_{\ro{CMB}}$ for which the contour lines of $(\gamma v)_{\ro{CMB}}$ are seen to bend correspond to the threshold~$ F_{\ro{BR}}$, which we showed in Figure~\ref{fig:feq}.
In the lower regions of each plot where 
$F_{\mathrm{CMB}} < F_{\mathrm{BR}}$, the backreaction on the IGMFs is negligible and we have $v_{\mathrm{CMB}} = v_0$, which is  independent of the MM density. 
On the other hand in the upper regions where $F_{\mathrm{CMB}} > F_{\mathrm{BR}}$, the backreaction is significant and $v_{\mathrm{CMB}} = v_{\ro{max}}$; here the MM velocity depends on the flux but is independent of the IGMF coherence length, as shown in \eqref{eq:gamma_v_max}.
The dashed gray line in the plot shows where $v_{\mathrm{CMB}} = 10^{-3}$. 
This value is the same order as the peculiar velocity of the Milky Way, which will become important in the next section.

In Figure~\ref{fig:MonAcc} we show $(\gamma v)_{\ro{CMB}}$ as a function of the MM mass for fixed values of the flux:
$F_{\mathrm{CMB}} = 10^{-30}\, \mathrm{cm}^{-2} \, \mathrm{sr}^{-1}\,  \mathrm{s}^{-1}$ (solid lines) and
$10^{-20}\, \mathrm{cm}^{-2} \, \mathrm{sr}^{-1} \, \mathrm{s}^{-1}$ (dashed).
The IGMF strength and coherence length are taken as 
$B_{\mathrm{I}} = 10^{-15}\, \mathrm{G}$
and $\lambda_{\mathrm{I}}=1\, \mathrm{Mpc}$ for the red lines,
while $B_{\mathrm{I}} = 10^{-10}\, \mathrm{G}$
and $\lambda_{\mathrm{I}} > 1/H_0$ for the blue lines.
The flux on the solid lines satisfy $F_{\mathrm{CMB}} < F_{\mathrm{BR}}$,
and thus the MM velocity is set by $v_0$. On the other hand, the dashed lines have $F_{\mathrm{CMB}} > F_{\mathrm{BR}}$ and the MMs move at~$v_{\mathrm{max}}$.

The above results show that, due to the acceleration in IGMFs, 
intermediate to low mass MMs can become relativistic. This remains true even when taking into account the backreaction on the IGMFs, albeit with a reduction in the MM velocity.

\section{Modification of Parker bound}
\label{sec:Parker}

\subsection{Parker bound}

The abundance of MMs within the Milky Way Galaxy is constrained by the requirement that the MMs should not short out the Galactic magnetic fields~\cite{Parker:1970xv,Turner:1982ag}. 
The upper bound on the MM flux inside the Milky Way thus obtained is commonly referred to as the Parker bound, and takes the form~\cite{Kobayashi:2023ryr}
 \begin{equation}
\label{eq:parker}
\begin{aligned}
 F_{\ro{MW}} \lesssim \ro{max.} \Biggl\{
&10^{-16}\, \mathrm{cm}^{-2} \mathrm{sec}^{-1} \mathrm{sr}^{-1} \left(\frac{m \left(\gamma_{\mathrm{MW}} - 1 \right)}{10^{11}\, \mathrm{GeV}} \right) \left( \frac{g}{g_{\mathrm{D}}} \right)^{-2} 
\left( \frac{\lambda_{\mathrm{G}}}{1 \, \mathrm{kpc}} \right)^{-1}
\left(\frac{\tau_{\mathrm{gen}}}{10^{8}\, \mathrm{yr}} \right)^{-1},
\\
 &10^{-16}\, \mathrm{cm}^{-2} \mathrm{sec}^{-1} \mathrm{sr}^{-1} \left( \frac{g}{g_{\mathrm{D}}} \right)^{-1}
\left( \frac{B_{\mathrm{G}}}{10^{-6}\, \mathrm{G}} \right)
\left( \frac{R}{\lambda_{\mathrm{G}}} \right)^{1/2}
\left(\frac{\tau_{\mathrm{gen}}}{10^{8}\, \mathrm{yr}} \right)^{-1}
\Biggr\} .
\end{aligned}
\end{equation}
Here $\gamma_{\mathrm{MW}}$ is the Lorentz factor of the MMs in the rest frame of the Milky Way, $\lambda_{\mathrm{G}}$ is the coherence length of the Galactic magnetic field, $R$ is the size of the magnetic field region of the Galaxy,
$\tau_{\mathrm{gen}}$ is the dynamo time scale for the amplification of the Galactic fields,
and $B_{\mathrm{G}}$ is the mean Galactic field strength.\footnote{The parameters including~$\gamma_{\ro{MW}}$ need to satisfy a few conditions
for the Parker bound to hold; see Eqs.~(2.16)-(2.18) in~\cite{Kobayashi:2023ryr}. 
These are satisfied for the parameter choices in this paper.}
One can further require the survival of the initial seed magnetic field of the Galaxy, the flux bound from which takes the same form except for that $B_{\mathrm{G}}$ is taken as the seed field strength~\cite{Adams:1993fj}.\footnote{We note that the seed Parker bound of the form~(\ref{eq:parker}) neglects the cosmic expansion since the initial time when the dynamo begins to operate. 
The bound can be modified for seed fields at high redshifts.} 

The Parker bound applies to both MMs that are clustered with the Galaxy, and unclustered MMs which only pass through the Galaxy.
However, with the typical parameters of the Galactic field shown as the reference values in the above equation, 
MMs with a Dirac charge can stay clustered until today only if they are ultraheavy as $ m \gtrsim 10^{18}\, \ro{GeV}$~\cite{Turner:1982ag,Kobayashi:2023ryr}. 
As we are primarily interested in intermediate to low mass MMs which are significantly accelerated in IGMFs, in the following we focus on unclustered MMs.

The Parker bound uses the entire Galaxy as a MM detector, and consequently the bound depends on the MMs' incident velocity on the Galaxy. 
The first line of~(\ref{eq:parker}) actually depends on~$\gamma_{\ro{MW}}$, and in particular, the Parker bound becomes weaker for larger~$\gamma_{\ro{MW}}$. 
This is because fast-moving MMs pass through the Galaxy while being minimally deflected by the Galactic field,
hence do not effectively dissipate the magnetic energy.
In the literature, the incident velocity has always been assumed to be
$v_{\ro{MW}} \sim 10^{-3}$
(i.e. $\gamma_{\ro{MW}} - 1 \sim 10^{-6}$),
as both the Milky Way's peculiar velocity and the MMs' gravitational infall velocity are of this order.\footnote{The same value is used for clustered MMs since the virial velocity of the Milky Way is also of this order.}
However we have seen in the previous section that MMs can be accelerated to much larger velocities in the IGMFs.
In the following, we study how the presence of IGMFs affect the Parker bound.

\subsection{Monopole velocity in the Milky Way's rest frame}

The peculiar velocity of the Milky Way Galaxy with respect to the CMB rest frame, $v_{\ro{p}} \sim 10^{-3}$, sources a relative velocity between the cosmic MMs and the Galaxy.
However if the MM velocity~$v_{\ro{CMB}}$ obtained in the IGMFs is even larger, this would dominate the relative velocity.
Therefore, the MM velocity in the rest frame of the Milky Way can be written as
\begin{equation}
 v_{\ro{MW}} = \ro{max.} \left\{ v_{\ro{p}}, v_{\ro{CMB}} \right\}.
\label{eq:v_MW}
\end{equation}
We also denote the incident flux of MMs on the Milky Way by $F_{\ro{MW}} = n v_{\ro{MW}} / 4 \pi$;
this is equivalent to the MM flux inside the Milky Way from the conservation of the number of MMs.
In Figure~\ref{fig:MonAccCMB} we showed the combination of $m$ and $F_{\ro{CMB}}$ for which $v_{\mathrm{CMB}} \sim v_{\mathrm{p}}$ by the dashed gray line. 
On the left of the gray line where $v_{\ro{CMB}} > v_{\ro{p}}$, 
the quantities $v_{\ro{CMB}}$ and $F_{\ro{CMB}}$ are identical to $v_{\ro{MW}}$ and $F_{\ro{MW}}$, respectively. 
On the other hand, on the right of the gray line where $v_{\ro{CMB}} < v_{\ro{p}}$, 
the MM velocity in the Milky Way's rest frame becomes~$v_{\ro{p}}$. 
Therefore, contour plots of $(\gamma v)_{\ro{MW}}$ in the 
$m$-$F_{\ro{MW}}$~plane are the same as in Figure~\ref{fig:MonAccCMB} except for that on the right of the gray dashed lines, 
$(\gamma v)_{\ro{MW}}$ is fixed to a constant value of~$(\gamma v)_{\ro{p}}$.

\subsection{Parker bound in the presence of intergalactic magnetic fields}

\begin{figure}[h!t!] 
    \includegraphics[width=0.65\textwidth]{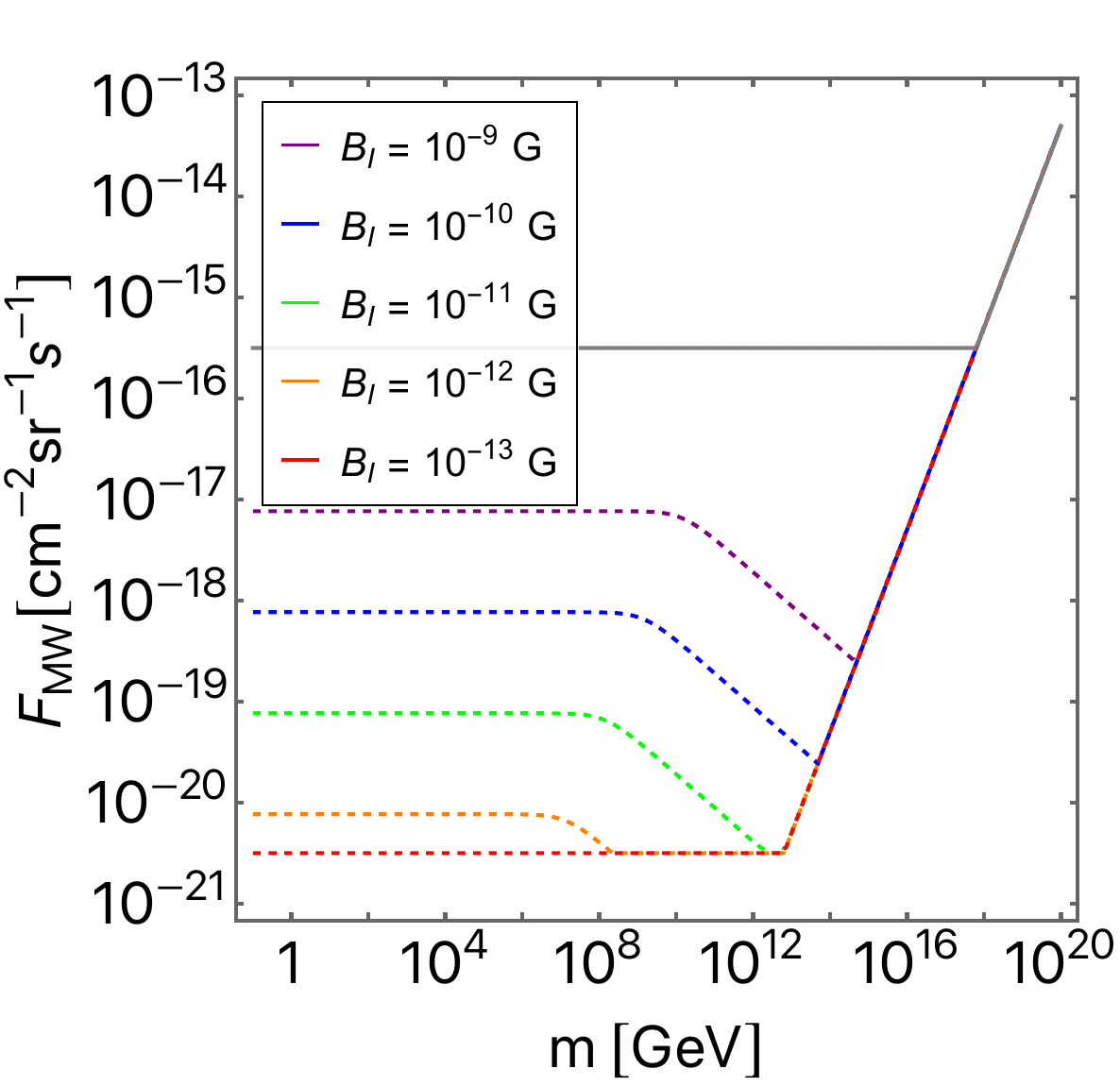}
    \centering
    \caption{Galactic Parker bound and seed Galactic Parker bound in the presence of IGMFs with strength $B_{\mathrm{I}} \leq 10^{-9}\, \ro{G}$. 
The MM charge is fixed to $g = g_{\mathrm{D}}$.
The Galactic Parker bound is not affected by the IGMFs and thus is shown by a single gray solid line. 
The seed Galactic Parker bound is affected, which is shown by dashed curves with different colors corresponding to different IGMF strengths.
The Galactic field strength is taken as $B_{\mathrm{G}} = 10^{-6}\, \mathrm{G}$, and the seed field as $B_{\mathrm{G}} = 10^{-11}\, \mathrm{G}$, with the other parameters taken as 
$\lambda_{\mathrm{G}} = 1 \, \mathrm{kpc}$, $R = 10 \, \mathrm{kpc}$, $\tau_{\mathrm{gen}} = 10^{8}\, \mathrm{yr}$.
The curves overlap in the right part of the plot.}
    \label{fig:Parker}
\end{figure}

In Figure~\ref{fig:Parker} we plot the Parker bound in Eq.~\eqref{eq:parker}, with the MM Lorentz factor~$\gamma_{\ro{MW}}$ evaluated through Eq.~(\ref{eq:v_MW}). 
The MM charge is fixed to $g = g_{\mathrm{D}}$.
The gray solid line shows the bound from the survival of
Galactic fields with $B_{\mathrm{G}} = 10^{-6}\, \mathrm{G}$,
while the dashed curves are from seed fields with an assumed strength of $B_{\mathrm{G}} = 10^{-11}\, \mathrm{G}$. 
The other Galactic parameters are fixed to 
$\lambda_{\mathrm{G}} = 1 \, \mathrm{kpc}$, $R = 10 \, \mathrm{kpc}$, $\tau_{\mathrm{gen}} = 10^{8}\, \mathrm{yr}$, and
$v_{\ro{p}} = 10^{-3}$.
The IGMF strength is varied between 
$B_{\ro{I}} = 10^{-9}\, \ro{G}$ and $B_{\ro{I}} = 10^{-13}\, \ro{G}$.
IGMFs within this range do not affect the Galactic bound, which hence is shown by a single solid line. 
On the other hand, the seed bound is strongly affected by the presence of IGMFs; dashed lines of different colors correspond to different IGMF strengths as shown in the plot legend. 
The Parker bounds for a fixed MM velocity of
$v_{\ro{MW}} = 10^{-3}$, 
as have been assumed in the literature, 
are reproduced for sufficiently weak IGMFs.
The Galactic and seed bounds in the weak IGMF limit are shown respectively by the gray solid and red dashed curves.
All the curves join at large masses, in other words, 
IGMFs modify the seed bound at intermediate to low masses.
We also note that with the choice of parameters, the flux bounds are independent of the IGMF correlation length,
as long as it is as large as $\lambda_{\ro{I}} \gtrsim 10^{-5}\, \ro{Mpc}$.
This is because for values of $F_{\ro{MW}}$ that saturate the bound, $v_{\ro{MW}}$ is set either by $v_{\ro{p}}$ or $v_{\ro{max}}$, both of which do not depend on~$\lambda_{\ro{I}}$. 
This can also be seen from Figures~\ref{fig:Parker} and \ref{fig:feq} showing that the upper bounds of $F_{\ro{MW}}$ are larger than the threshold value~$F_{\ro{BR}}$ for the backreaction to the IGMFs to become relevant.

For MMs with large kinetic energies, 
the expression~(\ref{eq:parker}) for the Parker bound is set by the first line which explicitly depends on the kinetic energy.
On the other hand for small kinetic energies, the bound is set by the second line which depends neither on the MM mass nor on the velocity.
The presence of IGMFs thus modify the Parker bound if the MMs are accelerated to velocities larger than the Milky Way's peculiar velocity, and further if the MM kinetic energy is sufficiently large such that the 
first line of the Parker bound applies.

For the choice of parameters in the plot, 
the upper bound on the flux 
from Galactic fields ($B_{G} = 10^{-6} \, \ro{G}$) 
in the weak IGMF limit is of 
$F_{\ro{MW}} \sim 10^{-16}\, \mathrm{cm}^{-2} \mathrm{sec}^{-1}  \mathrm{sr}^{-1}$ or larger.
The threshold kinetic energy between the two expressions for the Parker bound is $m (\gamma_{\ro{MW}} - 1) \sim 10^{11}\, \ro{GeV}$.
For flux values saturating the bound,
the energy that MMs acquire from IGMFs with $B_I \leq 10^{-9}\, \ro{G}$ is smaller than the threshold, which is why the Galactic bound is not altered by IGMFs. 
On the other hand, seed fields ($B_{G} = 10^{-11} \, \ro{G}$) give an upper bound in the weak IGMF limit of
$F_{\ro{MW}} \sim 10^{-21}\, \mathrm{cm}^{-2} \mathrm{sec}^{-1} \mathrm{sr}^{-1}$ or larger,
with the threshold energy between the two expressions being 
$m (\gamma_{\ro{MW}} - 1) \sim 10^{6}\, \ro{GeV}$.
For these flux values, the energy can exceed the threshold with IGMFs of 
$B_{\ro{I}} \gtrsim 10^{-12}\, \ro{G}$, hence the seed Parker bound is modified.

In order to understand the behavior of the modified Parker bound,
note that substituting Eq.~(\ref{eq:v_MW}) for $\gamma_{\ro{MW}}$ into the first line of Eq.~\eqref{eq:parker}
yields an upper limit which itself depends on the MM flux.
In particular when $v_{\ro{MW}} = v_{\ro{max}}$
(i.e. $v_{\ro{p}} < v_{\ro{max}}  < v_0$), 
one sees from Eq.~(\ref{eq:gamma_v_max}) that
$(\gamma v)_{\ro{MW}} \propto 1/ \left( m F_{\ro{MW}} \right)$.
For sufficiently small masses such that $v_{\ro{MW}}$ is relativistic, 
the kinetic energy scales as $ m \gamma_{\ro{MW}} \propto 1/F_{\ro{MW}}$.
Thus the first line of Eq.~\eqref{eq:parker} yields an inequality of the form 
$F_{\ro{MW}} < \kappa / F_{\ro{MW}}$ where $\kappa$ is a factor that is independent of the MM mass and abundance. Solving this gives a flux bound
$F_{\ro{MW}} < \kappa^{1/2}$, which is independent of the MM mass, as shown in the plot for the modified seed Parker bound at small masses.
On the other hand for larger masses with nonrelativistic~$v_{\ro{MW}}$, 
the kinetic energy scales as
$ m v_{\ro{MW}}^2/2 \propto 1/m F_{\ro{MW}}^2$,
and hence the first line of Eq.~\eqref{eq:parker} takes a form 
$F_{\ro{MW}} < \kappa / m F_{\ro{MW}}^2$.
This gives a bound $F_{\ro{MW}} < (\kappa / m)^{1/3}$, which is stronger for larger masses, as is seen for the middle parts of the modified seed bounds in the plot.

Before ending this section, we should remark that if a sufficiently strong IGMF existed at the time of Galaxy formation, 
this could have grown during a protogalactic collapse into the currently observed Galactic field. 
In this scenario, the necessity for a seed Galactic field and the dynamo amplification is obviated, and hence the Parker bounds are modified; 
in particular, the seed Parker bound is nullified.
Our findings indicate that even if a seed field coexisted with a strong IGMF and hence the seed Parker bound holds, it would be significantly relaxed due to the acceleration effect.

\section{Conclusion}
\label{sec:concl}

In light of the modern understanding of IGMFs, we have comprehensively analyzed the acceleration of MMs in the intergalactic space.
We found in particular that MMs with intermediate to low masses are accelerated to relativistic velocities. The results are shown in Figure~\ref{fig:MonAccCMB}.

The MM acceleration in IGMFs has significant effects on direct and indirect searches. As one such example, we studied how the Galactic Parker bounds are modified for fast-moving MMs.
Although it has been assumed in the literature that the initial MM velocity with respect to the Milky Way Galaxy is of the order of the peculiar velocity or the virial velocity of the Galaxy, this is not necessarily the case in the presence of IGMFs.
We showed that the large MM velocity weakens the so-called extended Parker bound based on the survival of seed Galactic magnetic fields. 
On the other hand, the Parker bound from the present-day Galactic field is unaffected by IGMFs that are compatible with observations.

Any additional magnetic fields around the Milky Way, such as those transported by galactic winds or outflow-driven bubbles, can also accelerate the MMs and further weaken the Parker bound. 
An accurate study of the effects of such fields requires numerical analyses, which we leave for future work.

We also showed that if the cosmic flux of MMs
is larger than the threshold values of Eqs.~\eqref{eq:feq_homo} or~\eqref{eq:feq_nonhomo} (which are also illustrated in Figure~\ref{fig:feq}),
their backreaction on IGMFs is non-negligible.
In this case the energy oscillates between the IGMFs and the MMs, and the oscillation-averaged MM velocity depends on the MM density.
Future developments in studies of IGMFs may be able to constrain such IGMF-MM oscillations, which will allow us to improve bounds on the MM flux.

This paper serves as a first step towards a complete understanding of MM acceleration in cosmic environments.
An accurate knowledge of the velocity of MMs entering the Earth 
is crucial for interpreting the constraints from terrestrial detectors 
(e.g. \cite{IceCube:2021eye,PierreAuger:2016imq})
in terms of the MM mass. 
The acceleration in IGMFs can outweigh the acceleration by Galactic fields, as we showed in Appendix~\ref{app:MonAccTot}.
We will discuss the implications 
for MM experiments in a companion work~\cite{Perri:inprep}.
It would also be interesting to study relativistic MMs as a possible explanation of the recently claimed detection of an extremely high-energy cosmic ray~\cite{TelescopeArray:2023sbd};
see also the earlier papers~\cite{porter1960dirac,Kephart:1995bi,Escobar:1997mr,Berezinsky:1998ft,Berezinsky:1999ec} which studied similar ideas.
Finally, we note that a cosmic population of relativistic MMs from IGMFs can serve as dark radiation, which may be further constrained from cosmological studies.

\acknowledgments
We thank Ivan De Mitri, 
Pranjal Ralegankar, Piero Ullio, and Matteo Viel for helpful discussions.
The work of T.K. was supported by the European Union - NextGenerationEU, in the framework of the PRIN Project ``Charting unexplored avenues in Dark Matter'' (20224JR28W). 
T.K. also acknowledges support from INFN TAsP and JSPS KAKENHI (JP22K03595). 
MD acknowledges support by the Italian MUR Departments of Excellence grant 2023-2027 “Quantum Frontiers.”

\appendix

\section{Primordial contribution to the monopole velocity}
\label{app:acceleration}

In the main text, we did not specify the origin of the IGMFs. 
In particular in Section~\ref{sec:acc}, we evaluated the velocity of MMs that have been accelerated in IGMFs for a Hubble time. 
However it is often claimed in the literature that IGMFs have a primordial origin.\footnote{See, e.g., \cite{Gruzinov:2001ah} for an alternative explanation.}
Based on this premise, one might inquire about the relevance of 
MM acceleration in primordial magnetic fields, since primordial fields would have been much stronger at higher redshifts. 
In this appendix, we demonstrate that the acceleration in primordial fields only give an order-unity correction to the results derived 
in Section~\ref{sec:acc}.

Let us consider for simplicity a homogeneous primordial magnetic field. The equation of motion of a MM in an expanding universe filled with a homogeneous magnetic field with strength~$B$
is~\cite{Kobayashi:2022qpl,Kobayashi:2023ryr}
\begin{equation}
    m \frac{d}{dt} (\gamma v) = g B - f_{\ro{p}} v - m H \gamma v,
    \label{eq:evolution}
\end{equation}
where we ignored velocity components perpendicular to the direction of the magnetic field. 
The last two terms in the right-hand side denote frictional forces from 
the MM scattering on charged particles in the primordial plasma, and from the cosmological expansion, with $H$ being the Hubble rate.

Before electron-positron annihilation, one can write the friction coefficient~$f_{\ro{p}}$ as~\cite{Long:2015cza,Kobayashi:2022qpl}
\begin{equation}
    f_p \sim \frac{e^2 g^2 \mathcal{N}_c}{16 \pi^2} T^2,
\end{equation}
where $\mathcal{N}_c$ is the number of relativistic and electrically charged degrees of freedom in thermal equilibrium, and $T$ is the temperature of the plasma. 
Considering MMs with charges not too far from the Dirac charge, 
the Hubble friction is negligible compared to the friction from the plasma~\cite{Kobayashi:2022qpl,Kobayashi:2023ryr}.
The MMs thus approach a terminal velocity which results from equating the first two terms in the right-hand side of Eq.~\eqref{eq:evolution} as
\begin{equation}
\label{eq:terminal}
    v_{\text{p}} = \frac{gB}{f_{\ro{p}}} \sim 10^{-8} \left(\frac{g_D}{g}\right)
    \left(\frac{B_{\rm I}}{10^{-15} G}\right).
\end{equation}
We note that the terminal velocity before the electron-positron annihilation is constant, since  
the homogeneous magnetic field~$B$ and the friction coefficient~$f_{\ro{p}} \propto T^2$ both redshift\footnote{Magnetic fields with small coherence lengths can evolve differently, when the 
wave mode is inside the Hubble horizon~\cite{Durrer:2013pga}.}
as $(1+z)^2$.
In the far right-hand side of Eq.~(\ref{eq:terminal}), 
$B_{\ro{I}}$ represents the present-day magnetic field strength. 

After electron-positron annihilation, the number of the free charged particles drops by $10$ orders of magnitude, so one can neglect the
plasma friction, i.e. the second term in the right-hand side of Eq.~(\ref{eq:evolution}). 
Then the present-day MM velocity can be obtained by integrating Eq.~\eqref{eq:evolution} as
\begin{equation}
    (\gamma v)_0 = \frac{(\gamma v)_{\rm p}}{1 + z_{\rm an}} + \int\limits_0^{z_{\text{an}}} \frac{g B(z)}{m H(z) (1+z)^2} dz
\approx \frac{(\gamma v)_{\rm p}}{1 + z_{\rm an}} +
 3 \times \frac{g B_{\mathrm{I}}}{m H_0},
   \label{eq:gammavPrim}
\end{equation}
where $z_{\text{an}} \sim 10^9$ is the redshift at electron-positron annihilation. 
Since the homogeneous magnetic field redshifts as $B(z) \propto (1+z)^2$, the only redshift dependence in the integrand comes from the Hubble rate $H(z)$.
Upon moving to the far right-hand side, 
we have used the cosmological parameters for $\Lambda$CDM cosmology~\cite{Planck:2018vyg} and numerically integrated the second term.
Using Eq.~(\ref{eq:terminal}), the ratio between the two terms in 
Eq.~(\ref{eq:gammavPrim}) is written as
\begin{equation}
\frac{(\gamma v)_{\rm p}}{1 + z_{\rm an}}
\left( \frac{g B_{\mathrm{I}}}{m H_0} \right)^{-1}
\sim 10^{-8} \left( \frac{g_{\ro{D}}}{g} \right)^2
 \left( \frac{m}{10^{18}\, \ro{GeV}} \right),
\end{equation}
which shows that the first term is negligible for MM masses considered in this paper. 
Hence we obtain the velocity of MMs accelerated in a homogeneous primordial magnetic field as
\begin{equation}
 (\gamma v)_0  \approx 3 \times \frac{g B_{\mathrm{I}}}{m H_0}.
\end{equation}
This matches with the result given in Eq.~(\ref{eq:homo}) at the order-of-magnitude level. 
We therefore conclude that the order-of-magnitude results of this paper remain unaffected, even if the IGMFs have a primordial origin.

\section{Energy loss of fast monopoles in a nonrelativistic plasma}
\label{app:energy_loss_medium}

A MM traveling in a plasma loses its kinetic energy as it scatters off electrically charged particles. 
We evaluate the drag force experienced by the MM by following the procedure outlined in Section~14.2 of \cite{Vilenkin:2000jqa}.
However here we consider the plasma to consist of nonrelativistic particles, and focus on cases where the MM velocity is much larger than the velocities of the individual plasma particles in the rest frame of the plasma.

We start in the rest frame of a MM. The magnetic field around the MM with charge~$g$ is given by $\bd{B} = g \bd{r} / 4 \pi r^3$.
(In this appendix $g$ is not necessarily the charge amplitude and thus can be negative.) 
Hence the equation of motion of a particle with electric charge~$q$,
velocity $\bd{u}$, and momentum~$\bd{p}$ is
\begin{equation}
 \frac{d\bd{p}}{dt} = \frac{qg}{4 \pi }
\frac{\bd{u} \times \bd{r}}{r^3}.
\end{equation}
Integrating this yields the total change in the momentum of the particle scattered by the MM's magnetic field.
Focusing on small-angle scatterings, and hence ignoring the effect of the magnetic field on the particle's trajectory,
one obtains
\begin{equation}
 \Delta p = \frac{\abs{qg}}{4 \pi }
\int^{\infty}_{- \infty} dt
\frac{u \sin \phi }{r^2}
= \frac{\abs{qg}}{ 2 \pi b},
\end{equation}
where $\phi$ is the angle between $\bd{r}$ and $\bd{u}$, 
and $b$ is the impact parameter.
Upon moving to the far right-hand side we used 
$\sin \phi = b / r$, and 
$r = \sqrt{ (ut)^2 + b^2}$ with $t = 0$ corresponding to when the distance between the particle and the MM is minimized.
The scattering angle is $\theta = \Delta p / p$, with which the differential cross section is obtained as
\begin{equation}
 \frac{d \sigma }{d \theta } 
= - 2 \pi b \frac{db}{d \theta }
= \frac{(q g)^2}{2 \pi p^2 \theta^3}.
\end{equation}

We assume that in the rest frame of the plasma, the MM velocity~$\bd{v}$ has a much larger amplitude compared to the velocities of the individual plasma particles.\footnote{A large relative velocity between the MM and the plasma particles also supports the assumption of an unperturbed particle trajectory.}
Then in the MM rest frame, the particles can be considered to move with a universal velocity $\bd{u} = - \bd{v}$,
and the scatterings with a collection of the particles 
induce the MM to experience a force,
\begin{equation}\label{eq:maru-5}
\begin{split}
  \bd{F} &= \gamma n_{\ro{p}} \int d \theta \frac{d \sigma }{d \theta }
u \bd{p} (1- \cos \theta ) 
\\
& = \frac{C (qg)^2 n_{\ro{p}}}{4 \pi m_{\ro{p}}}
\frac{\bd{u}}{u}.
\end{split}
\end{equation}
Here $m_{\ro{p}}$ represents the mass of the plasma particles, and $n_{\ro{p}}$ is the particle number density in the plasma rest frame;
note that in the MM rest frame the number density is enhanced by the Lorentz factor $\gamma = 1 / \sqrt{1 - v^2}$.
The particle momentum is written as 
$\bd{p} = m_{\ro{p}} \gamma \bd{u}$,
and we used that the average momentum transfer for a scattering with a small angle~$\theta$ is $\bd{p} (1 - \cos \theta)$. 
Upon moving to the second line, we have written the integral over small angles as
\begin{equation}
 C = 2 \int d\theta \frac{1- \cos \theta }{\theta^3} 
\simeq \int \frac{d \theta }{\theta }
= \ln \frac{\theta_{\ro{max}}}{\theta_{\ro{min}}}.
\end{equation}
The upper limit of integration~$\theta_{\ro{max}}$ can be set to unity as we are focusing on small-angle scatterings. 
The lower limit~$\theta_{\ro{min}}$ arises from the fact that the above calculation based on single-particle scatterings breaks down if the 
impact parameter~$b$ becomes larger than the mean free path of the particles in the plasma and/or the Debye length.
Here we do not evaluate the detailed value of~$\theta_{\ro{min}}$ since $\bd{F}$ is only logarithmically sensitive to it. 

The expression for the drag force can be covariantized by noting that the 
force vector~$\mathcal{F}^\mu$ should, in the MM rest frame, have 
spatial components matching with (\ref{eq:maru-5}),
along with a vanishing time component since the MM's energy does not change.
These are satisfied by a force vector with the form,
\begin{equation}
 \mathcal{F}^{\mu} = - 
\frac{C (qg)^2 n_{\ro{p}}}{4 \pi m_{\ro{p}} }
\frac{1}{v_{\ro{rel}} }
\left\{
\frac{\ro{u}^{\mu} }{\ro{u}_{\nu} \ro{v}^{\nu}}
+ \ro{v}^{\mu}
\right\},
\end{equation}
where we use $(-+++)$ for the metric signature, 
$\ro{v^\mu}$ is the four-velocity of the MM, 
$\ro{u^\mu}$ is the four-velocity of the plasma,
and $v_{\ro{rel}} = \sqrt{1 - ( \ro{u}_\nu \ro{v}^\nu)^{-2}}$ denotes the amplitude of the relative velocity between the MM and the plasma
(in the MM rest frame, 
the spatial components of $\ro{u}^\mu$ reduce to $\gamma \bd{u}$, and 
$v_{\ro{rel}} = u$.)

The covariant equation of motion of a MM with mass~$m$ 
in electromagnetic fields is given by
\begin{equation}
 m \ro{v}^\nu \nabla_{\nu} \ro{v}^{\mu} = 
g \tilde{F}^{\mu \nu } \ro{v}_{\nu} + \mathcal{F}^{\mu},
\end{equation}
with $\tilde{F}^{\mu \nu}$ being the dual electromagnetic field tensor. 
Taking a Minkowski metric and supposing that in the rest frame of the plasma there is only magnetic fields but no electric fields, then the equation of motion in the plasma rest frame reduces to 
\begin{equation}
 m \frac{d}{dt} (\gamma \bd{v}) = g \bd{B} - 
\frac{C (qg)^2 n_{\ro{p}}}{4 \pi m_{\ro{p}}} \frac{\bd{v}}{v}.
\label{eq:maru-7}
\end{equation}

Let us now compare the deceleration 
and acceleration terms in (\ref{eq:maru-7}) in cosmic environments, by estimating their ratio:
\begin{equation}
 \epsilon = \frac{1}{g B} \frac{C (qg)^2 n_{\ro{p}}}{4 \pi m_{\ro{p}}}.
\end{equation}
As particles with smaller masses yield a larger drag force, 
here we focus on free electrons (instead of protons).
In the intergalactic space,
using the electron density in the intergalactic medium (IGM)
$n_e \sim 1 \, \ro{m}^{-3}$,
the IGMF lower limit $B_{\ro{I}} \sim 10^{-15}\, \ro{G}$,
and further taking 
$g = g_{\ro{D}}$ and\footnote{Even if $C$ is larger by a few orders of magnitude, our main conclusion does not change.}
$C \sim 1$, 
one finds $\epsilon \sim 10^{-10}$.
In the Milky Way, considering instead the interstellar medium (ISM) with 
$n_e \sim 1 \, \ro{cm}^{-3}$
and Galactic fields of $B_{\ro{G}} \sim 10^{-6}\, \ro{G}$ gives
$\epsilon \sim 10^{-13}$.
We thus see that both in the intergalactic space and the Milky Way, 
since the electron density is so small such that $n_{e} / m_{e} \ll B $, 
the acceleration of MMs in the magnetic fields completely dominates over the deceleration by scattering free electrons.

Let us comment on effects that we have not taken into account.
Firstly, the above calculation is based on classical single-particle scatterings. 
A quantum field theory computation may give rise to corrections to the scattering cross section, especially in the ultrarelativistic regime. 
We also note that the picture of single-particle scatterings breaks down at distances larger than the Debye length, where plasma effects become important. 
These have been studied in \cite{Hamilton:1983qme,Meyer-Vernet:1985yyn} for slow MMs, for which it was shown that the energy loss from plasma effects is typically not significantly larger than that from single-particle scatterings.
Finally, we remark that MMs can further loose energy through 
ionization of neutral particles, atomic excitations, bremsstrahlung, 
electron-pair production, and photonuclear interactions;
the last two processes become particularly important in the ultrarelativistic regime. 
(See e.g. \cite{Wick:2000yc} for a study of various energy-loss processes for MMs.
A rough estimate of the interactions with neutral particles in the ISM is given in \cite{Kobayashi:2023ryr}, according to which they are negligible for a wide range of parameters.)
It would be important to study all these effects, in particular in the ultrarelativistic regime. We leave this for future work.

Before ending this appendix, we also estimate the energy loss of MMs through the emission of electromagnetic radiation.
Using the Li\'enard formula (i.e. relativistic version of the Larmor formula),
and exchanging the electric charge to magnetic,
the energy loss per time of an accelerating MM is 
\begin{equation}
P \sim  g^2 \gamma^6 \left\{
(\dot{\bd{v}})^2 - (\bd{v} \times \dot{\bd{v}})^2
\right\}
= \frac{g^4 B^2}{m^2},
\end{equation}
where an overdot denotes a time-derivative.
Upon moving to the far right-hand side, we have considered a MM accelerating along a homogeneous magnetic field and dropped the cross product, and also used the equation of motion~(\ref{eq:maru-7}) without the drag force term.
The ratio between the radiative energy loss, and energy gain per time from the acceleration in the magnetic field, $g B v$ (here we take $g > 0$), is thus
\begin{equation}
 \delta \sim \frac{g^3 B}{m^2 v}.
\end{equation}
For IGMFs of $B_{\ro{I}} = 10^{-15}\, \ro{G}$, relativistic ($ v \sim 1$) MMs with $g = g_{\ro{D}}$ give 
$\delta \sim 10^{-31} (1\, \ro{GeV} / m)^2$.
For Galactic fields of $B_{\ro{G}} = 10^{-6}\, \ro{G}$,
this becomes $\delta \sim 10^{-22} (1\, \ro{GeV} / m)^2$.
In either case, radiative emissions are negligible for 
relativistic MMs as long as the mass satisfies $m^2 \gg g^3 B$.

\section{Monopole-magnetic field oscillation}
\label{app:oscillation}

If MMs accelerated in magnetic fields do not dissipate their kinetic energy into the ambient plasma, they eventually return the energy to the magnetic fields.
In this way the energy can move back and forth between the magnetic field and monopoles (which would be the magnetic analog of the Langmuir oscillation \cite{PhysRev.33.195}). However one would expect such oscillations to be subject to Landau damping, if the phase velocity of the magnetic field were smaller than the random velocity of the individual MMs~\cite{Turner:1982ag}.

In order to estimate the phase velocity, we start by considering a homogeneous one-dimensional system where the magnetic field has a component~$B$ along a certain direction, and the MMs with charge~$g$~($>0$) have a velocity component~$v$ along the same direction.
(In this appendix, $B$ and $v$ are not the amplitudes and thus can be negative.)
Then by combining the MMs' equation of motion,
$m \, d (\gamma v) / dt  = g B$,
with the conservation of total energy density
$\rho_{\ro{tot}} = n m \gamma + B^2 / 2$
where $n$ is the MM number density, 
one obtains an evolution equation for the magnetic field,
\begin{equation}
 \frac{d^2 B}{dt^2} + \Omega^2 B = 0.
\end{equation}
Here $\Omega$ is the effective frequency taking the form:
\begin{equation}
\Omega =  \frac{g}{\gamma^{3/2}} \left(\frac{n}{m} \right)^{1/2}.
\label{eq:Omega}
\end{equation}
This becomes time-independent for nonrelativistic MMs ($\gamma \simeq 1$), hence let us focus on this case for simplicity.
Further supposing that inhomogeneous fields with a finite coherence length~$\lambda$ also oscillate with this frequency, then the magnetic field's phase velocity is obtained as
$ v_{\ro{ph}} = \lambda \Omega / 2 \pi $.

\textit{Galactic magnetic fields.}
The phase velocity of Galactic fields with coherence length~$\lambda_{\ro{G}}$ is of
\begin{equation}
 v_{\ro{ph}} \sim 
10^{-5}
\left( \frac{v_{\ro{MW}}}{10^{-3}} \right)^{-1/2}
\left( \frac{F_{\ro{MW}}}{10^{-15} \, \ro{cm}^{-2} \ro{sec}^{-1} \ro{sr}^{-1}} \right)^{1/2}
 \left(\frac{g}{g_{\ro{D}}} \right)
\left( \frac{m}{10^{18}\, \ro{GeV}} \right)^{-1/2}
\left( \frac{\lambda_{\ro{G}}}{1\, \ro{kpc}} \right).
\end{equation}
We have rewritten $n$ in terms of the MM flux in the rest frame of the Milky Way Galaxy (cf. below (\ref{eq:v_MW}); however here we consider both clustered and unclustered monopoles). The reference value for the flux is set to the Galactic Parker limit at $m = 10^{18}\, \ro{GeV}$
(cf. Figure~\ref{fig:Parker}).
Note that MMs with a Dirac charge and mass $m \gtrsim 10^{18}\, \ro{GeV}$ can cluster with the Galaxy~\cite{Turner:1982ag,Kobayashi:2023ryr}.
Such clustered MMs obtain a virial velocity of~$10^{-3}$, which is larger than the magnetic field's phase velocity;
hence the oscillation is expected to rapidly evaporate. 
The phase velocity can in principle become larger for smaller masses, 
however light MMs do not cluster and hence should pass through the Galaxy before (completely) returning the energy to the Galactic fields.
A flux much larger than the Parker limit can also increase the phase velocity beyond the virial velocity, however \cite{Parker:1987} claimed that oscillating Galactic fields do not match with observations,
and also that the inhomogeneity in the MM distribution further leads to the damping of the oscillations.
For these reasons, one can trust the Parker bound without worrying about the possibility that the Galactic fields survive as oscillating fields.\footnote{Primordial magnetic fields can also be used to obtain Parker-type bounds, because MMs effectively dissipate energy into the primordial plasma and thus energy oscillations do not happen even without Landau damping~\cite{Long:2015cza,Kobayashi:2022qpl,Kobayashi:2023ryr}.
This is in contrast to the tiny energy loss of MMs in the present-day IGM or ISM, which we showed in Appendix~\ref{app:energy_loss_medium}.}

\textit{Intergalactic magnetic fields.}
The phase velocity of IGMFs is of
\begin{equation}
 v_{\ro{ph}} \sim 
10^{-5} \, v_{\ro{CMB}}^{-1/2}
\left( \frac{F_{\ro{CMB}}}{10^{-27} \, \ro{cm}^{-2} \ro{sec}^{-1} \ro{sr}^{-1}} \right)^{1/2}
 \left(\frac{g}{g_{\ro{D}}} \right)
\left( \frac{m}{10^{10}\, \ro{GeV}} \right)^{-1/2}
\left( \frac{\lambda_{\ro{I}}}{1\, \ro{Mpc}} \right).
\end{equation}
Here we rewrote $n$ in terms of the flux in the CMB rest frame (cf. below (\ref{eq:v_CMB})),
and for its reference value used the flux threshold where the MM backreaction becomes relevant 
for IGMFs with $B_{\ro{I}} = 10^{-15}\, \ro{G}$ at $m = 10^{10}\, \ro{GeV}$ (cf. Figure~\ref{fig:feq}).
Note that the phase velocity increases for smaller~$m$ and $v_{\ro{CMB}}$.
MMs in the intergalactic space can obtain random velocities from scattering with the IGM and/or from gravitational potentials. 
However we expect these to be much smaller than the IGMF's phase velocity,
and hence in this paper we consider the IGMF-MM oscillations to survive.

\section{Monopole acceleration in intergalactic and galactic magnetic fields}
\label{app:MonAccTot}

\begin{figure*}[h!t!]
     \centering
     \hspace{35pt}
     \begin{subfigure}[b]{0.55\textwidth}
         \centering
         \includegraphics[width=0.8\textwidth]{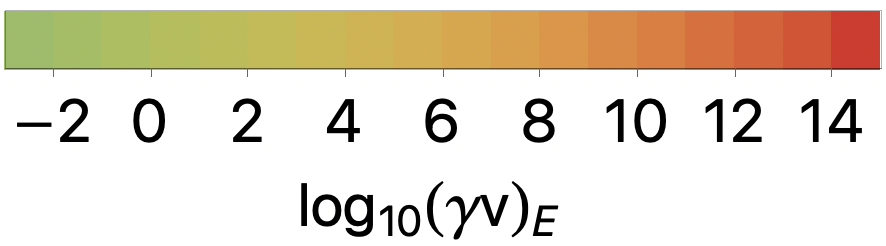}
     \end{subfigure}
     \hspace{65pt}
     \begin{subfigure}[b]{0.493\textwidth}
         \centering
         \includegraphics[width=\textwidth]{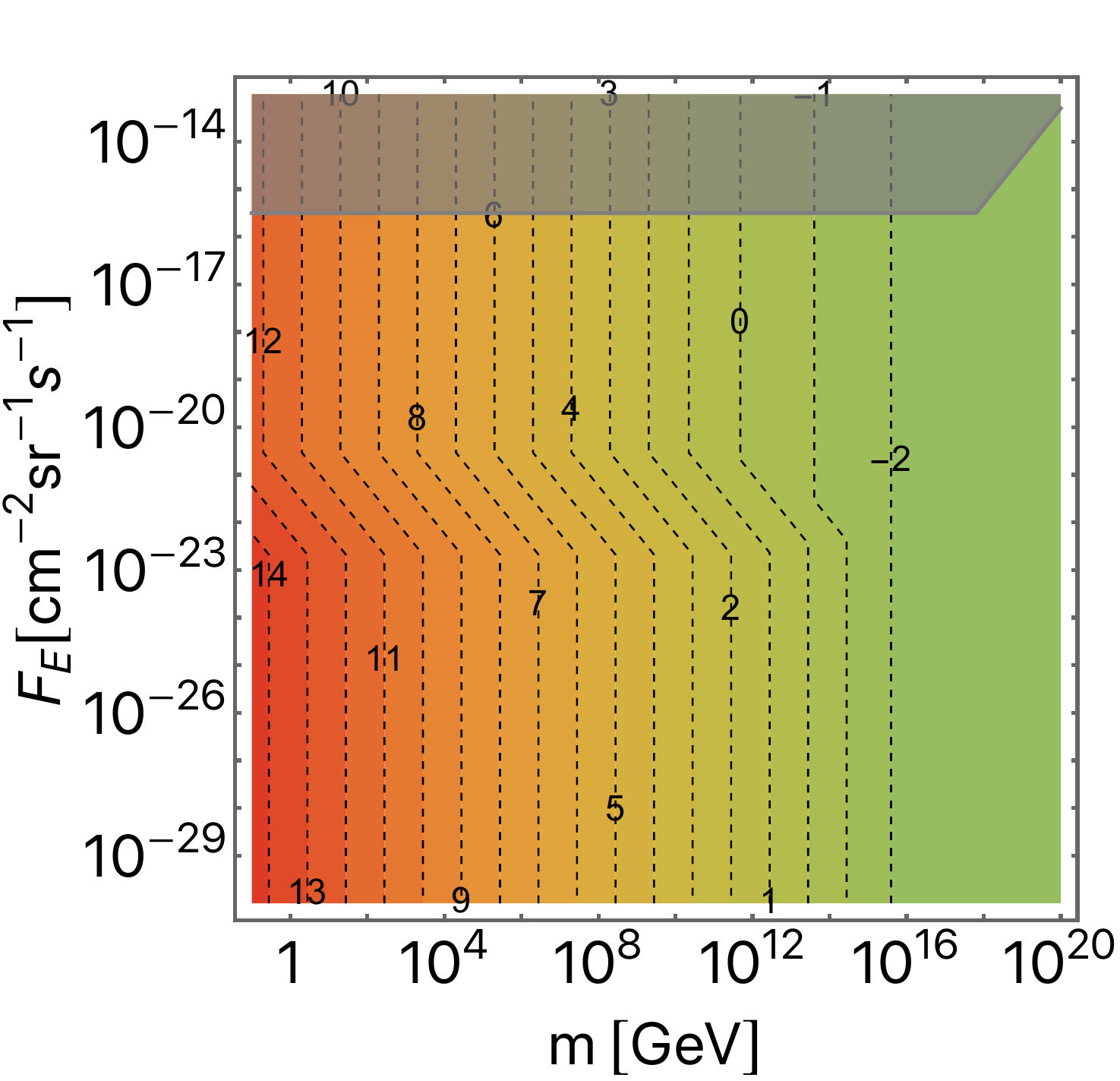}
         \caption{$B_{\mathrm{I}} = 10^{-10}\, \ro{G}$.}
         \label{fig:MW10}
     \end{subfigure}
     \hfill
     \begin{subfigure}[b]{0.493\textwidth}
         \centering
         \includegraphics[width=\textwidth]{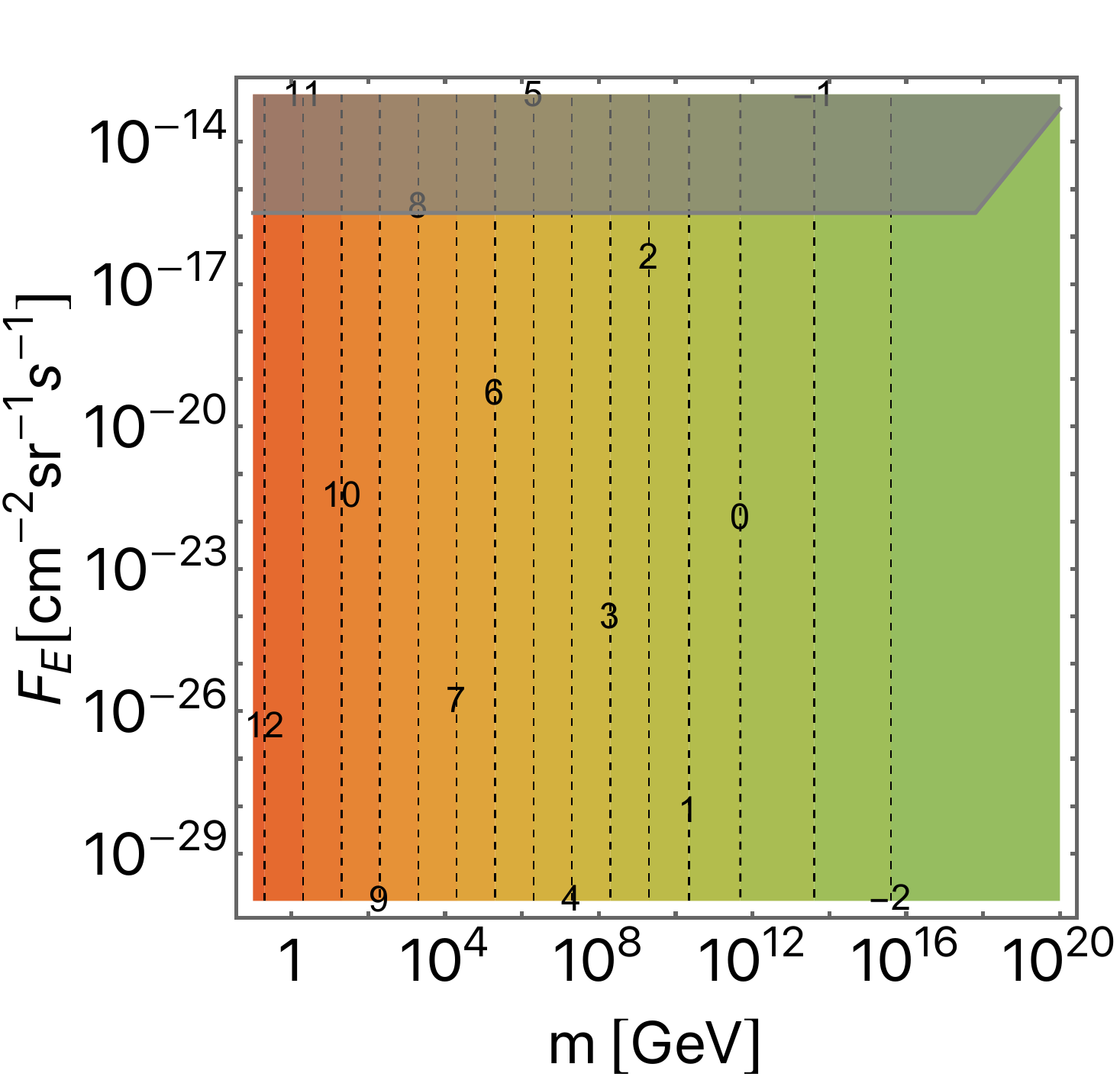}
         \caption{$B_{\mathrm{I}} = 10^{-15}\, \ro{G}$.}
         \label{fig:MW15}
     \end{subfigure}
     \caption{MM velocity on Earth evaluated by taking into account the acceleration in intergalactic and Galactic magnetic fields. 
The contours show $\log_{10} ( \gamma v )_{\mathrm{E}}$.
The parameters for the Galactic magnetic field are fixed to $B_{\mathrm{G}} = 10^{-6} \, \mathrm{G}$, $\lambda_{\mathrm{G}} = 1\, \mathrm{kpc}$, and $R = 10 \, \mathrm{kpc}$. The IGMF is taken as homogeneous ($\lambda_{\mathrm{I}} > 1/H_0$), with the strength varied in each panel. The MM charge is fixed to $g = g_{\mathrm{D}}$. The color scheme is the same as in Figure~\ref{fig:MonAccCMB}. The gray region is excluded by the Galactic Parker bound.}
        \label{fig:MonAccMW}
\end{figure*}

In this appendix we study the velocity of MMs arriving on Earth, by focusing on MMs that are not clustered with the Milky Way. Such unclustered MMs are first accelerated in the IGMFs, then after entering the Milky Way they are further accelerated in the Galactic magnetic fields.

Let us suppose for simplicity that MMs cross the magnetic field region of the Milky Way before arriving at Earth. 
Hence in terms of the size of the magnetic region~$R$
and the Galactic field's coherence length~$\lambda_{\mathrm{G}}$,
the MMs pass through $N \sim R / \lambda_{\ro{G}}$ number of cells of uniform magnetic field.
Then the typical kinetic energy that each MM obtains from Galactic fields is estimated using an expression similar to~(\ref{eq:B3}) as
\begin{equation}
 m (\gamma_{\ro{G}} - 1 ) \sim g B_{\ro{G}} \sqrt{R \lambda_{\ro{G}}},
\label{eq:C.1}
\end{equation}
with $B_{\ro{G}}$ being the Galactic field strength.
If this is larger than the MM's initial kinetic energy upon entering the Milky Way, 
$m (\gamma_{\ro{MW}} - 1)$,
then the acceleration in Galactic fields predominates.
Therefore, the velocity of MMs at Earth in the Milky Way's rest frame
can be written as
\begin{equation}
\label{eq:vEarth}
    v_{\mathrm{E}} = \ro{max.} \left\{v_{\mathrm{G}}, v_{\ro{MW}} \right\}
= \ro{max.} \left\{v_{\mathrm{G}}, v_{\ro{p}}, v_{\ro{CMB}} \right\},
\end{equation}
where we used (\ref{eq:v_MW}), and
$\gamma_{\ro{G}} = (1-v_{\ro{G}}^2)^{-1/2}$.
We note that the velocity of the Earth with respect to the Milky Way is $\sim 10^{-3}$, which is comparable to the Milky Way's peculiar velocity~$v_{\ro{p}}$.
Since $v_{\ro{E}} \geq v_{\ro{p}}$
as is clear from~(\ref{eq:vEarth}), the MM velocity in the rest frame of the Milky Way matches with that in the rest frame of Earth, at least at the order-of-magnitude level.
Hence $v_{\ro{E}}$ can also be considered as the relative velocity between the MMs and Earth.\footnote{(\ref{eq:C.1}) implicitly assumes that there is enough time for MMs to reach Earth solely by the acceleration in Galactic fields. This assumption can break down if, for instance, $B_{\ro{G}}$ is tiny.
However in such cases $v_{\ro{E}}$ is equivalent to~$v_{\ro{MW}}$,
and one can still use (\ref{eq:vEarth}) combined with (\ref{eq:C.1}) to estimate the MM velocity at Earth.}
For the same reason, and also because the flux of unclustered MMs is conserved inside the Galaxy, the MM flux incident on the Milky Way~$F_{\ro{MW}}$ is the same as that on Earth~$F_{\ro{E}}$, i.e. $F_{\ro{MW}} = F_{\ro{E}}$.

In Figure~\ref{fig:MonAccMW} we show the values of $v_{\mathrm{E}}$ in \eqref{eq:vEarth} as a function of the MM mass and flux on Earth. 
Here we assume a homogeneous IGMF with its amplitude varied as
$B_{\mathrm{I}} = 10^{-10} \, \mathrm{G}$ (left) and
$10^{-15} \, \mathrm{G}$ (right),
while the parameters for the Galactic magnetic field are taken as $B_{\mathrm{G}} = 10^{-6}\, \mathrm{G}$, $\lambda_{\mathrm{G}} = 1 \, \mathrm{kpc}$, and $R = 10 \, \mathrm{kpc}$. 
The MM charge is fixed to $g = g_{\mathrm{D}}$,
and the color scheme follows that in Figure~\ref{fig:MonAccCMB}.
The left plot exhibits a variety of behaviors of the MM velocity. 
In the upper region, the acceleration by Galactic fields is dominant over that by IGMFs; here the velocity is set by $v_{\mathrm{G}}$ and does not depend on the MM flux. 
In the central region, the acceleration by IGMFs dominates with the MMs' backreaction being significant; hence the velocity is given by~$v_{\ro{max}}$.
In the lower region, the acceleration by IGMFs dominates with negligible backreaction; the velocity is set by~$v_{0}$ which is independent of the MM flux. 
In the right part, the velocity is set by the peculiar velocity of the Milky Way, $v_{\mathrm{p}} \sim 10^{-3}$.
On the other hand in the right plot with a weaker IGMF, the acceleration from Galactic fields completely dominates over that from IGMFs for all values of the MM flux. 
From (\ref{eq:C.1}), the mass dependence of the velocity induced by the Galactic field can be read off as
$(\gamma v)_{\ro{G}} \propto m^{-1}$ at small masses where $v_{\ro{G}} \simeq 1$, while 
$(\gamma v)_{\ro{G}} \propto m^{-1/2}$ at large masses where $v_{\ro{G}} \ll 1$;
these behaviors are actually seen in the plots.

Note that upon evaluating the acceleration in~(\ref{eq:C.1}), we have ignored the backreaction of the MMs on the Galactic fields. 
In reality, the MM flux $F_{\ro{E}}$ ($ = F_{\ro{MW}}$) cannot exceed the Galactic Parker bound shown in Figure~\ref{fig:Parker}, otherwise the MMs would short out the Galactic fields and contradict with observations.
In the plots, the regions excluded by the Galactic Parker bound are shown in gray.

Hence we have demonstrated that the MM velocity on Earth can be governed either by the acceleration in IGMFs, in Galactic fields, or by the Milky Way's peculiar velocity. Which of them dominates depends on a number of parameters, including the MM flux, mass, and the relative strength of the intergalactic and Galactic fields.

\bibliographystyle{unsrt}
\bibliography{biblio}

\providecommand{\noopsort}[1]{}\providecommand{\singleletter}[1]{#1}%
\begin{thebibliography}{10}

\bibitem{Dirac:1931kp}
Paul Adrien~Maurice Dirac.
\newblock {Quantised singularities in the electromagnetic field,}.
\newblock {\em Proc. Roy. Soc. Lond. A}, 133(821):60--72, 1931.

\bibitem{tHooft:1974kcl}
Gerard 't~Hooft.
\newblock {Magnetic Monopoles in Unified Gauge Theories}.
\newblock {\em Nucl. Phys. B}, 79:276--284, 1974.

\bibitem{Polyakov:1974ek}
Alexander~M. Polyakov.
\newblock {Particle Spectrum in Quantum Field Theory}.
\newblock {\em JETP Lett.}, 20:194--195, 1974.

\bibitem{Zeldovich:1978wj}
Ya.~B. Zeldovich and M.~Yu. Khlopov.
\newblock {On the Concentration of Relic Magnetic Monopoles in the Universe}.
\newblock {\em Phys. Lett. B}, 79:239--241, 1978.

\bibitem{Preskill:1979zi}
John Preskill.
\newblock {Cosmological Production of Superheavy Magnetic Monopoles}.
\newblock {\em Phys. Rev. Lett.}, 43:1365, 1979.

\bibitem{Tavecchio:2010mk}
F.~Tavecchio, G.~Ghisellini, L.~Foschini, G.~Bonnoli, G.~Ghirlanda, and P.~Coppi.
\newblock {The intergalactic magnetic field constrained by Fermi/LAT observations of the TeV blazar 1ES 0229+200}.
\newblock {\em Mon. Not. Roy. Astron. Soc.}, 406:L70--L74, 2010.

\bibitem{Neronov:2010gir}
A.~Neronov and I.~Vovk.
\newblock {Evidence for strong extragalactic magnetic fields from Fermi observations of TeV blazars}.
\newblock {\em Science}, 328:73--75, 2010.

\bibitem{Dermer:2010mm}
Charles~D. Dermer, Massimo Cavadini, Soebur Razzaque, Justin~D. Finke, James Chiang, and Benoit Lott.
\newblock {Time Delay of Cascade Radiation for TeV Blazars and the Measurement of the Intergalactic Magnetic Field}.
\newblock {\em Astrophys. J. Lett.}, 733:L21, 2011.

\bibitem{MAGIC:2022piy}
V.~A. Acciari et~al.
\newblock {A lower bound on intergalactic magnetic fields from time variability of 1ES 0229+200 from MAGIC and Fermi/LAT observations}.
\newblock {\em Astron. Astrophys.}, 670:A145, 2023.

\bibitem{Durrer:2013pga}
Ruth Durrer and Andrii Neronov.
\newblock {Cosmological Magnetic Fields: Their Generation, Evolution and Observation}.
\newblock {\em Astron. Astrophys. Rev.}, 21:62, 2013.

\bibitem{AlvesBatista:2021sln}
Rafael Alves~Batista and Andrey Saveliev.
\newblock {The Gamma-ray Window to Intergalactic Magnetism}.
\newblock {\em Universe}, 7(7):223, 2021.

\bibitem{Broderick:2011av}
Avery~E. Broderick, Philip Chang, and Christoph Pfrommer.
\newblock {The Cosmological Impact of Luminous TeV Blazars I: Implications of Plasma Instabilities for the Intergalactic Magnetic Field and Extragalactic Gamma-Ray Background}.
\newblock {\em Astrophys. J.}, 752:22, 2012.

\bibitem{Miniati:2012ge}
Francesco Miniati and Andrii Elyiv.
\newblock {Relaxation of Blazar Induced Pair Beams in Cosmic Voids: Measurement of Magnetic Field in Voids and Thermal History of the IGM}.
\newblock {\em Astrophys. J.}, 770:54, 2013.

\bibitem{Perry:2021rgv}
Roy Perry and Yuri Lyubarsky.
\newblock {The role of resonant plasma instabilities in the evolution of blazar induced pair beams}.
\newblock {\em Mon. Not. Roy. Astron. Soc.}, 503(2):2215--2228, 2021.

\bibitem{Alawashra:2022all}
Mahmoud Alawashra and Martin Pohl.
\newblock {Suppression of the TeV Pair-beam\textendash{}Plasma Instability by a Tangled Weak Intergalactic Magnetic Field}.
\newblock {\em Astrophys. J.}, 929(1):67, 2022.

\bibitem{Kephart:1995bi}
Thomas~W. Kephart and Thomas~J. Weiler.
\newblock {Magnetic monopoles as the highest energy cosmic ray primaries}.
\newblock {\em Astropart. Phys.}, 4:271--279, 1996.

\bibitem{Escobar:1997mr}
C.~O. Escobar and R.~A. Vazquez.
\newblock {Are high-energy cosmic rays magnetic monopoles?}
\newblock {\em Astropart. Phys.}, 10:197--202, 1999.

\bibitem{Wick:2000yc}
Stuart~D. Wick, Thomas~W. Kephart, Thomas~J. Weiler, and Peter~L. Biermann.
\newblock {Signatures for a cosmic flux of magnetic monopoles}.
\newblock {\em Astropart. Phys.}, 18:663--687, 2003.

\bibitem{Akahori:2017lhe}
Takuya Akahori et~al.
\newblock {Cosmic Magnetism in Centimeter and Meter Wavelength Radio Astronomy}.
\newblock {\em Publ. Astron. Soc. Jap.}, 70(1):R2, 2018.

\bibitem{Garcia:2020kxm}
Andres~Aramburo Garcia, Kyrylo Bondarenko, Alexey Boyarsky, Dylan Nelson, Annalisa Pillepich, and Anastasia Sokolenko.
\newblock {Magnetization of the intergalactic medium in the IllustrisTNG simulations: the importance of extended, outflow-driven bubbles}.
\newblock {\em Mon. Not. Roy. Astron. Soc.}, 505(4):5038--5057, 2021.

\bibitem{Erceg_2022}
Ana Erceg, Vibor Jelić, Marijke Haverkorn, Andrea Bracco, Timothy~W. Shimwell, Cyril Tasse, John~M. Dickey, Lana Ceraj, Alexander Drabent, Martin~J. Hardcastle, and Luka Turić.
\newblock Faraday tomography of lotss-dr2 data: I. faraday moments in the high-latitude outer galaxy and revealing loop iii in polarisation.
\newblock {\em Astronomy and Astrophysics}, 663:A7, 2022.

\bibitem{Carretti:2022fqk}
E.~Carretti, S.~P. O\textquoteright{}Sullivan, V.~Vacca, F.~Vazza, C.~Gheller, T.~Vernstrom, and A.~Bonafede.
\newblock {Magnetic field evolution in cosmic filaments with LOFAR data}.
\newblock {\em Mon. Not. Roy. Astron. Soc.}, 518(2):2273--2286, 2022.

\bibitem{OSullivan:2023eub}
S.~P. O'Sullivan et~al.
\newblock {The Faraday Rotation Measure Grid of the LOFAR Two-metre Sky Survey: Data Release 2}.
\newblock {\em Mon. Not. Roy. Astron. Soc.}, 519(4):5723--5742, 2023.

\bibitem{Heesen_2023}
V.~{Heesen}, S.~P. {O'Sullivan}, M.~{Br{\"u}ggen}, A.~{Basu}, R.~{Beck}, A.~{Seta}, E.~{Carretti}, M.~G.~H. {Krause}, M.~{Haverkorn}, S.~{Hutschenreuter}, A.~{Bracco}, M.~{Stein}, D.~J. {Bomans}, R.~J. {Dettmar}, K.~T. {Chy{\.z}y}, G.~H. {Heald}, R.~{Paladino}, and C.~{Horellou}.
\newblock {Detection of magnetic fields in the circumgalactic medium of nearby galaxies using Faraday rotation}.
\newblock {\em Astronomy and Astrophysics}, 670:L23, 2023.

\bibitem{Bertone:2006mr}
Serena Bertone, Corina Vogt, and Torsten Ensslin.
\newblock {Magnetic Field Seeding by Galactic Winds}.
\newblock {\em Mon. Not. Roy. Astron. Soc.}, 370:319--330, 2006.

\bibitem{Parker:1970xv}
Eugene~N. Parker.
\newblock {The Origin of Magnetic Fields}.
\newblock {\em Astrophys. J.}, 160:383, 1970.

\bibitem{Turner:1982ag}
Michael~S. Turner, Eugene~N. Parker, and T.~J. Bogdan.
\newblock {Magnetic Monopoles and the Survival of Galactic Magnetic Fields}.
\newblock {\em Phys. Rev. D}, 26:1296, 1982.

\bibitem{Adams:1993fj}
Fred~C. Adams, Marco Fatuzzo, Katherine Freese, Gregory Tarle, Richard Watkins, and Michael~S. Turner.
\newblock {Extension of the Parker bound on the flux of magnetic monopoles}.
\newblock {\em Phys. Rev. Lett.}, 70:2511--2514, 1993.

\bibitem{Rephaeli:1982nv}
Yoel Rephaeli and Michael~S. Turner.
\newblock {The Magnetic Monopole Flux and the Survival of Intracluster Magnetic Fields}.
\newblock {\em Phys. Lett. B}, 121:115--118, 1983.

\bibitem{Grasso:2000wj}
Dario Grasso and Hector~R. Rubinstein.
\newblock {Magnetic fields in the early universe}.
\newblock {\em Phys. Rept.}, 348:163--266, 2001.

\bibitem{Subramanian:2015lua}
Kandaswamy Subramanian.
\newblock {The origin, evolution and signatures of primordial magnetic fields}.
\newblock {\em Rept. Prog. Phys.}, 79(7):076901, 2016.

\bibitem{Long:2015cza}
Andrew~J. Long and Tanmay Vachaspati.
\newblock {Implications of a Primordial Magnetic Field for Magnetic Monopoles, Axions, and Dirac Neutrinos}.
\newblock {\em Phys. Rev. D}, 91:103522, 2015.

\bibitem{Kobayashi:2022qpl}
Takeshi Kobayashi and Daniele Perri.
\newblock {Parker bound and monopole pair production from primordial magnetic fields}.
\newblock {\em Phys. Rev. D}, 106(6):063016, 2022.

\bibitem{Kobayashi:2023ryr}
Takeshi Kobayashi and Daniele Perri.
\newblock {Parker bounds on monopoles with arbitrary charge from galactic and primordial magnetic fields}.
\newblock {\em Phys. Rev. D}, 108(8):083005, 2023.

\bibitem{Kogut:1993ag}
A.~Kogut et~al.
\newblock {Dipole anisotropy in the COBE DMR first year sky maps}.
\newblock {\em Astrophys. J.}, 419:1, 1993.

\bibitem{MoEDAL:2021vix}
B.~Acharya et~al.
\newblock {Search for magnetic monopoles produced via the Schwinger mechanism}.
\newblock {\em Nature}, 602(7895):63--67, 2022.

\bibitem{IceCube:2021eye}
R.~Abbasi et~al.
\newblock {Search for Relativistic Magnetic Monopoles with Eight Years of IceCube Data}.
\newblock {\em Phys. Rev. Lett.}, 128(5):051101, 2022.

\bibitem{PierreAuger:2016imq}
Alexander Aab et~al.
\newblock {Search for ultrarelativistic magnetic monopoles with the Pierre Auger Observatory}.
\newblock {\em Phys. Rev. D}, 94(8):082002, 2016.

\bibitem{Perri:inprep}
Daniele Perri, Michele Doro, and Takeshi Kobayashi.
\newblock in preparation.

\bibitem{TelescopeArray:2023sbd}
R.~U. Abbasi et~al.
\newblock {An extremely energetic cosmic ray observed by a surface detector array}.
\newblock {\em Science}, 382:903--907, 2023.

\bibitem{porter1960dirac}
N.~A. Porter.
\newblock The dirac monopole as a constituent of primary cosmic radiation.
\newblock {\em Nuovo Cimento}, 16:958, 1960.

\bibitem{Berezinsky:1998ft}
Veniamin Berezinsky, Pasquale Blasi, and Alexander Vilenkin.
\newblock {Ultrahigh-energy gamma-rays as signature of topological defects}.
\newblock {\em Phys. Rev. D}, 58:103515, 1998.

\bibitem{Berezinsky:1999ec}
V.~Berezinsky.
\newblock {Ultrahigh-energy cosmic rays from cosmological relics}.
\newblock {\em Nucl. Phys. B Proc. Suppl.}, 87:387--396, 2000.

\bibitem{Gruzinov:2001ah}
Andrei Gruzinov.
\newblock {Grb phenomenology, shock dynamo, and the first magnetic fields}.
\newblock {\em Astrophys. J. Lett.}, 563:L15, 2001.

\bibitem{Planck:2018vyg}
N.~Aghanim et~al.
\newblock {Planck 2018 results. VI. Cosmological parameters}.
\newblock {\em Astron. Astrophys.}, 641:A6, 2020.
\newblock [Erratum: Astron.Astrophys. 652, C4 (2021)].

\bibitem{Vilenkin:2000jqa}
A.~Vilenkin and E.~P.~S. Shellard.
\newblock {\em {Cosmic Strings and Other Topological Defects}}.
\newblock Cambridge University Press, 7 2000.

\bibitem{Hamilton:1983qme}
A.~J.~S. Hamilton and C.~L. Sarazin.
\newblock {Deceleration of Grand Unified Theory monopoles in a plasma}.
\newblock {\em Astrophys. J.}, 274:399--407, 1983.

\bibitem{Meyer-Vernet:1985yyn}
N.~Meyer-Vernet.
\newblock {Energy loss by slow magnetic monopoles in a thermal plasma}.
\newblock {\em Astrophys. J.}, 290:21--23, 1985.

\bibitem{PhysRev.33.195}
Lewi Tonks and Irving Langmuir.
\newblock Oscillations in ionized gases.
\newblock {\em Phys. Rev.}, 33:195--210, Feb 1929.

\bibitem{Parker:1987}
Eugene~N. Parker.
\newblock {Magnetic Monopole Plasma Oscillations and the Survival of Galactic Magnetic Fields}.
\newblock {\em Astrophys. J.}, 321:349, 1987.

\end{thebibliography}

\end{document}